\documentclass[twocolumn,letterpaper]{aastex63}
\hypersetup{linkcolor=red,citecolor=blue,filecolor=cyan,urlcolor=magenta}
\usepackage{amsmath}
\usepackage{advdate}
\newcommand{\Mearth}{\mbox{$M_{\oplus}$}}            % Mearth
\newcommand{\Rearth}{\mbox{$R_{\oplus}$}}            % Rearth
\received{July 7, 2021}
\revised{February 15, 2022}
\accepted{March 10, 2022}
\submitjournal{The Planetary Science Journal}
\published{---}

\shorttitle{Formation of Composition Gradients Within Jupiter}
\shortauthors{Stevenson et al.}
\graphicspath{{./}{./}}
\begin{document}

\title{\uppercase{%
Mixing of Condensable Constituents with H-He During the Formation 
\& Evolution of Jupiter
}}

\correspondingauthor{David Stevenson}
\email{djs@gps.caltech.edu}

\author[0000-0001-9432-7159]{David J. Stevenson}
\affiliation{Division of Geological and Planetary Sciences, Caltech, Pasadena, CA 91125, USA}

\author[0000-0001-6093-3097]{Peter Bodenheimer}
\affiliation{UCO/Lick Observatory, Department of Astronomy and
Astrophysics, University of California, Santa Cruz, CA 95064, USA}
\email{peter@ucolick.org}

\author[0000-0001-6513-1659]{Jack J. Lissauer}
\affiliation{Space Science \& Astrobiology Division, 
MS 245-3,
NASA Ames Research Center, Moffett Field, CA 94035, USA}
\email{jack.lissauer@nasa.gov}

\author[0000-0002-2064-0801]{Gennaro D'Angelo}
\affiliation{Theoretical Division, Los Alamos National Laboratory, 
Los Alamos, NM 87545, USA}
\email{gennaro@lanl.gov}

%% Note that the \and command from previous versions of AASTeX is now
%% depreciated in this version as it is no longer necessary. AASTeX 
%% automatically takes care of all commas and "and"s between authors names.

%% AASTeX 6.3 has the new \collaboration and \nocollaboration commands to
%% provide the collaboration status of a group of authors. These commands 
%% can be used either before or after the list of corresponding authors. The
%% argument for \collaboration is the collaboration identifier. Authors are
%% encouraged to surround collaboration identifiers with ()s. The 
%% \nocollaboration command takes no argument and exists to indicate that
%% the nearby authors are not part of surrounding collaborations.

%% Mark off the abstract in the ``abstract'' environment. 
%% The abstract should be a single paragraph of not more than 250 words.
\begin{abstract}
Simulations of Jupiter’s formation are presented that incorporate mixing
of H-He with denser material entering the planet as solids.
Heavy compounds and gas mix substantially when the planet becomes roughly
as massive as Earth, because incoming planetesimals can fully vaporize.
Supersaturation of vaporized silicates causes the excess to sink as
droplets, but water remains at higher altitudes. Because the mean 
molecular weight decreases rapidly outward, some of the compositional
inhomogeneities produced during formation can survive for billions 
of years.
After $4.57$~Gyr, our Jupiter model retains compositional gradients;
proceeding outward one finds: i) an inner heavy-element core,
the outer part derived from hot supersaturated rain-out;
ii) a composition-gradient region, containing most of the heavy elements,
where H-He abundance increases outward, reaching about $0.9$ mass fraction
at $0.3$ of the radius, with silicates enhanced relative to water 
in the lower parts and depleted in the upper parts;
iii) a uniform composition region (neglecting He immiscibility) 
that is enriched over protosolar and contains most of the planet's mass;
and iv) an outer region where cloud formation (condensation)
of heavy constituents occurs.
This radial compositional profile has heavy elements more broadly
distributed than predicted by classical formation models, but 
less diluted than suggested by Juno-constrained gravity models. 
The compositional gradients in the region containing the bulk 
of the heavy elements prevent convection, in both our models
and those fitting current gravity, resulting in a hot interior 
where much of the accretion energy remains trapped.
\end{abstract}

%% Keywords should appear after the \end{abstract} command. 
%% See the online documentation for the full list of available subject
%% keywords and the rules for their use.

\section{Introduction} 
\label{sec:intro}

Any attempt to explain Jupiter should aspire to be compatible  
with the current structure, the main features of which have been partially 
clarified by the Juno mission and are described by \citet{stevenson2020}. 
Some aspects of that structure remain uncertain, but the key implication
is the presence of perhaps twenty or thirty Earth masses (\Mearth) of
heavy elements, denoted by $Z$ (the mass fraction of everything other than
hydrogen and helium), with some tendency toward a central concentration
of $Z$. Instead of the old well-defined core picture 
($Z(r) = 1$ for $r\le R_{c}$; $Z(r)=Z_{e}\ll 1$ for $r>R_{c}$; with 
radius $R_{c}\ll R_{\mathrm{J}}$, Jupiter's radius), new data 
suggest a ``dilute'' core \citep{wahl2017}, perhaps with a large stable 
region where $dZ/dr<0$. 
Ring seismology suggests Saturn may also have a dilute core, probably 
with an analogous stable region \citep{mankovich2021}.

The substantial (factor of five or more) average enrichment in $Z$ over
the solar composition must arise from some aspect of the formation process,
and the current distribution, $Z(r)$, is presumably also affected by initial 
conditions as well as by any processes of redistribution after formation. 
The emphasis on $Z(r)$ is appropriate because its contributions are likely 
to arrive as condensed matter that is partially or totally decoupled from 
the gaseous nebula in which they form. 

Re-assimilation of heavy elements into the gas phase within the planet 
(eventually a dense, hot fluid as the planet grows) can happen because 
of evaporation and mixing upon planetary assembly \citep{pollack1986}. 
Our paper focuses on the existence and consequences of this possible mixing. 
Although less susceptible to observation, it is also of interest to quantify 
the fate of the different $Z$ constituents: ice, silicates and iron. 
In particular, ice (H$_{2}$O) has very different physical 
behavior than silicates or iron. Interior models based on gravity
are insensitive to these distinctions, but the different contributions 
to $Z$ may be separable in deep atmospheric observations and might affect 
the convective history. 

Traditionally, planets were thought to separate into shells of iron,
silicates and ice because these materials are mutually immiscible at
relevant temperatures ($T$) and pressures ($P$), and have very different
densities. Current understanding implies otherwise: ice and rock can mix
in all proportions at the relevant $T$ and $P$ 
(A.\ Vazan, 2022, personal communication), and even silicates and iron may mix 
to the atomic level above $\approx 10\,000\,\mathrm{K}$. 
This precludes the old, sharply defined layering, but the
planetary growth simulations described in the present paper show that 
some differentiation can arise, at least between silicates and 
ices, simply because of the wildly different vaporization temperatures.
An incoming planetesimal that is a mixture of ice and rock
can break up and disseminate, with ice vaporizing
when its surface temperature is around $200$--$300\,\mathrm{K}$, 
while rock can only substantially vaporize for surface temperatures
around $2000$--$3000\,\mathrm{K}$. In both cases, the ambient
hydrogen-dominated atmosphere may oversaturate with the excess
condensable material raining out as liquid droplets, but with 
the vapor loading of the surrounding gas being much higher 
in ice than in rock high up, and the rock vapor loading increasing 
deeper down.
In both regions, $T$ and $P$ are low enough that mixing of ice 
and rock, thermodynamically favored at greater depths,
is not relevant. In short, ice and rock partially separate, 
not for the old reason (immiscibility) but because of the very 
different vaporization properties. The density differences are 
not directly relevant but become important in stabilization 
because the deeper vaporization occurs in the denser 
constituent. The result is a diffuse boundary between water 
and rock that can inhibit convection, with the observable 
atmosphere being enriched by the vaporized ices especially.  

A successful Jupiter model should also explain the observed
atmospheric abundances, the planetary heat flow and the magnetic field.
We are particularly concerned with the first of these here, since it may
be related to the assumed accretion history, or even to the delivery of
material from deep down to the outer layers of the planet. However,
our goal is not to compete with detailed models for the gravity
and magnetic fields, but rather to identify possible evolutionary 
stories leading to structures that are broadly compatible with data.
We note that heat flow is not a good way of deciding whether 
the interior is hotter than in fully adiabatic models, 
since most of the heat from Jupiter is emitted early and the current 
luminosity multiplied by the age of Jupiter is small relative 
to the heat content at the onset of the final cooling phase.

\subsection{Basic Formation Mechanisms}

In early work on Jupiter's formation, two main ideas were followed. 
The first assumed that Jupiter formation was analogous to the formation
of stars by gravitational (Jeans) instability 
\citep[see][and references therein]{boss2019}. This approach is not
currently favored and appears to have multiple problems, both for
the planets in our solar system and for the vast majority of known
exoplanets. 
The most frequently stated difficulty lies in the presence of $Z$
at abundances and spatial distributions that seem incompatible with
the simple Jeans picture; 
indeed this is why \citet{perri1974} abandoned it in favor of core
accretion, described below. There are also theoretical concerns that
center around the tendency for disks to redistribute material by
gravitational torques before the needed instability is reached.

We focus instead on the widely favored and more likely scenario
in which the formation of Jupiter (and giant planets in general)
is accomplished by first forming a core of heavy elements, 
embedded within a gaseous nebula, and followed by accumulation of gas.
The accretion of gas and solids is actually concurrent, but initially
solids dominate, whereas later gas is accreted at a much larger rate
than are solids.
This gas accretion is limited by the disk at some point, and eventually 
stopped by a combination of gap formation and the dispersal of the gaseous 
component of the protoplanetary disk.

Even in the core accretion scenario, there was still the notion of
a possible gravitational instability or equivalently a ``critical'' core
mass \citep{perri1974,mizuno1980,stevenson1982,bodenheimer1986,wuchterl1993}, 
but there was no fully consistent approach to the concurrent accretion 
of solids and gas until the seminal work of \citet{pollack1996}. 
They found that the accumulation of a massive envelope was mediated
by the need for the envelope to lose heat. An important consequence is
the absence of a well-defined critical mass, although it is still true 
that core mass and elapsed time determine the ability 
of the gas envelope to grow, so a core of order $10\,\Mearth$ is still
a relevant goal. 

\subsection{Formation Phases}

This led to a picture, still presumed to be largely correct, 
in which Jupiter's formation is divided into three phases, 
as identified in \citet{pollack1996}, here extended to five. 
In \textit{Phase~I}, solids accretion exceeds that of H-He,
hence the accreting body is high $Z$, except for the very low 
mass H-He gas envelope that surrounds it. 
This phase proceeds to a mass of at least several Earth masses. 
The surface density of solids within the nebula and the opacity 
of the forming envelope are important quantities. 
For realistic timescales of Phase~I ($10^{5}$--$10^{6}$ years), 
the gaseous envelope around this embryo is set by the assumed 
hydrostatic equilibrium from the embryo surface out to some 
fraction of the Hill or Bondi sphere (connecting to 
the surrounding nebula). 
In \textit{Phase~II}, when gas accretion is larger than solids 
accretion, the supply of $Z$ material is much diminished, because 
the zone providing accretion is largely depleted. 
Gas continues to accumulate, mediated by the cooling of 
the planet and the continued presence of a nearly hydrostatic
equilibrium extending out to the nebula. 
\textit{Phase~III} begins when the accreted mass of H-He gas 
equals the mass $M_{Z}$ within the planet, and the hydrostatic 
structure may be greatly affected by the very compressible 
gaseous envelope. For a gas-dominated planet like Jupiter, 
this phase ends when the planet's ability to acquire H-He 
exceeds what the surrounding disk can supply. 
The planet then enters \textit{Phase~IV} of disk-limited accretion
in which planet growth is dictated by disk thermodynamics. 
Such a limit was hypothesized by \citet{pollack1996}, 
but not included in formation models in a self-consistent 
manner until the work of \citet{lissauer2009}.
A surface shock may form at this stage as the planet 
detaches from the nebula.
The final mass, $M_{\mathrm{J}} \gg M_{Z}$, is reached 
in this phase and arises from processes that are not 
intrinsic to the planet, e.g., 
gap formation by disk-planet tidal interactions and 
the exhaustion of the accessible gas in the neighboring 
nebula. 
This general picture can be augmented by later accretion 
of solids, referred to here as \textit{Phase~V}, in order to explain 
the observed atmospheric enrichment of perhaps a factor 
of three over solar \citep{owen2003}.

\subsection{Objective of the Calculations}

The pioneering (and much subsequent) work on this picture omits 
consideration on the dissolution of solids in the envelope, and 
this is the first attempt to look in detail at how Jupiter’s structure 
is changed by a more realistic picture for the incoming $Z$ material. 
\citet{bodenheimer2018} consider many of the same issues, but for 
the simpler case in which ice is absent, motivated by the presence 
of light-gas envelopes of many exoplanets orbiting well interior to 
the ice condensation zone. 

This study forms a counterpart to the simplest extreme 
for Jupiter, which assumes that infalling solids reach
a condensed core, although all the energy of that infall 
is imparted to the gas. 
It is natural to suppose that we can thereby reconcile the ``dilute''
core (the current state of Jupiter) with formation models. 
Alas, it would appear that the story is not so simple because 
the models described here only go part way: They eliminate 
the sharply defined $Z=1$ core of almost all previous models 
of Jupiter's growth, but nonetheless produce a much more 
confined central region of nearly pure heavy elements than 
Jupiter's interior models seem to favor. 
This conundrum \citep[see also, e.g.,][]{mueller2020} 
does not at present have an explanation. 
Possible reconciliations include alternative accretion histories 
\citep[e.g.,][]{alibert2018}, a failure to consider fully 
the post-formation convective mixing, and the possible effects 
of ``giant impacts'' \citep[e.g.,][]{liu2019}.
Meteoritic evidence \citep{kruijer2017} 
is compatible with the general story of formation described here. 

\subsection{Miscibility and Stability}

These models ignore the late-stage onset of helium rain, the existence 
of which is not in doubt (because of Galileo probe measurements), and 
the consequences of which are mildly important for the modeling of current 
gravity. The motivations for this choice are twofold. 
First, it is not at all relevant 
to the epoch of planetary accretion that is the focus of this paper. 
Second, and perhaps less obviously, we know of no reason why the onset 
of helium rain would have major consequences for the distribution of $Z$ 
(even though we are aware of studies that presume this to be possible).

We expect that $Z(r)$ is related to the formation process and that 
incoming solids are reassimilated rather than reaching a $Z=1$ core 
intact for the following reasons. Once a significant mass (of order 
one Earth mass) is reached, incoming planetesimals can disaggregate 
into small particles because the ram pressure of the gas can exceed 
the compressive strength of the incoming body, much like for the Tunguska 
event in 1908. This material can rapidly vaporize, leading to a picture 
in which the mass of added solids relative to the concurring accretion 
of gas defines the mass fraction of heavy elements \citep{helled2017}.
Actually, our models suggest supersaturation and rain-out of excess 
condensate (silicate or water) as droplets, which reassimilate 
at greater depth as supercritical (single phase) fluid. 
The high temperature and increasing pressure ensure that there is 
no phase separation into immiscible layers, as is thermodynamically 
predicted for the formation of Earth \citep{stevenson1990}. 
In this sense, ``core'' formation in Jupiter is conceptually different 
from core formation in terrestrial planets, because the latter depends 
on immiscibility. At later times, when the relevant hydrogen is cooler 
but metallic, miscibility is predicted by quantum mechanics 
\citep{wilson2012}.

The higher density of $Z$ material relative to hydrogen is not relevant 
to the subsequent evolution except through its effect on convective vigor. 
In the presence of a pressure gradient (provided by gravity in this
instance), a static fluid of different constituents can separate diffusively 
(barodiffusion), but the consequences of this process
are small even for the age of the universe, for any mixture of interest. 
Accordingly, when we see that $dZ/dr\neq 0$ deep within in Jupiter, 
we must attribute this to something that happened in the delivery process 
and not to something that is ordained by mere consideration of thermodynamics 
and gravity. Moreover, a stable compositional gradient ($dZ/dr<0$) can have 
profound consequences for the thermal evolution of the body, especially 
if that body is mostly degenerate ($\alpha_{V} T \ll 1$, where $\alpha_{V}$
is the coefficient of thermal expansion).
The reason is that a modest compositional gradient can suppress convection.
For example, a doubling of the temperature deep down, say from 
$15\,000$ to $30\,000\,\mathrm{K}$, can be stabilized by increasing $Z$
by $0.2$. There are many factors that affect the formation of Jupiter,
and we focus on only one in this paper but, clearly, the magnitude and 
shape of $Z(r)$ are diagnostic, particularly since we seek to understand 
the origin of our solar system.

In the following, Section \ref{sec:methods} describes the computational 
methods and assumptions. Section~\ref{sec:results} gives the details 
on the results, and Section~\ref{sec:disc} discusses certain points 
regarding those results. A summary and conclusions are presented 
in Section \ref{sec:conclusions}.

\section{Methods} 
\label{sec:methods}

The basic procedure is the solution of the standard differential
equations of stellar structure in spherical symmetry: mass conservation, 
hydrostatic equilibrium, energy conservation, and energy transport 
by radiation, convection or conduction \citep{henyey1964,kippenhahn1990}. 
Extensive modifications of the physics to treat the planetary case include: 
\textit{i}) accounting for the  mass and energy deposition caused by
accretion 
of solids and gases from the primordial accretion disk; 
\textit{ii}) including equations of state of material rich in heavy
elements, such as water vapor and silicate vapor; 
\textit{iii}) inclusion of dust grains in the calculation of radiative
opacity; 
\textit{iv}) allowance for a central core composed entirely of heavy
elements; 
\textit{v}) accounting for the ablation and breakup of solid material 
(planetesimals) as it passes through the gaseous envelope of the planet;  
\textit{vi}) treatment of gradients in chemical composition that affect 
the convective transfer of material and energy; 
\textit{vii}) allowing for the effects of the central star in heating 
the planet and driving mass loss from it; and 
\textit{viii}) including aspects of the physics of the protoplanetary 
disk that affect accretion onto the planet.

The treatments of many of these effects are discussed in previous
papers.
\citet{pollack1996} and \citet{gennaro2014} summarize the interaction
of accreting planetesimals with the gaseous envelope, including ablation,
breakup, determination of the capture radius of the planet, and energy
deposition. \citet{bodenheimer2018} discuss the effects of composition 
gradients in the heavy elements, which can suppress the efficiency of 
convective transfer; they also summarize the main points regarding 
the equation of state and opacity, as well as the effects of the central
star. 
\citet{lissauer2009} describe three-dimensional hydrodynamic simulations 
of disk gas surrounding a forming planet. These simulations determine, 
first, an outer boundary condition on the planet, and, second, 
the maximum rate at which the disk can supply gas to the planet, 
as a function of planet mass, the ``unperturbed'' (in the absence 
of gravitational perturbations from the planet) surface mass density 
of the disk in the vicinity of the planet, and disk viscosity.
Once this limiting rate is reached, the planet contracts and hydrodynamic
flow onto it from the disk occurs.
The disk physics involved, and the accretion rates of gas and solids 
that are derived, are further discussed in \citet{bodenheimer2013} and 
\citet{gennaro2021}.

\subsection{Planetesimal Dissolution}
\label{sec:diss}

In the present work, Jupiter, forming at $5.2\,\mathrm{AU}$, accretes
planetesimals composed of water ice and rock (silicates, SiO$_{2}$),
assumed to have equal mass fractions. 
In \citet{pollack1996} and \citet{gennaro2014} the interaction
of the accreting planetesimals and the gaseous envelope is
described in detail. The trajectories of planetesimals
coming in at various impact parameters are integrated,
taking into account gravity and gas drag. Along the
trajectories, planetesimals lose mass by ablation, and
when the dynamical pressure on a planetesimal exceeds 
its compressive strength, it breaks up if the effects 
of self-gravity are overcome. 
The results from the various impact parameters are averaged 
to give the profiles of added mass and energy with radius. 
In the cited works, the deposited mass sinks to the core
(a well-defined entity under those assumptions), 
releasing gravitational energy on the way.
In contrast, in the present calculations mass and energy 
deposited by ablation and breakup can remain local.

As discussed in more detail in \citet{bodenheimer2018}, once vaporized 
at a given layer, the partial pressure $P_{\mathrm{part}}$ of a given 
substance is compared with the vapor pressure $P_{\mathrm{vap}}$ of
the substance. 
If $P_{\mathrm{part}} > P_{\mathrm{vap}}$, the excess material above
$P_{\mathrm{vap}}$ sinks (rains out) to a level where the temperature 
allows the two quantities to be equal. 
For water and silicates, respectively, the vapor pressures are
\begin{eqnarray}
P^{\mathrm{H_{2}O}}_{\mathrm{vap}} &=& 3.44 \times 10^{12} \exp{(-5640.34/T)} \label{eq:Pi_vap} \\ 
P^{\mathrm{SiO_{2}}}_{\mathrm{vap}} &=& 3.93 \times 10^{13} \exp{(-54700/T)},
\label{eq:Pr_vap}
\end{eqnarray}
where the temperature $T$ is in degrees kelvin and the pressure in
$\mathrm{dyne}/\mathrm{cm}^{2}$.
In Equations~(\ref{eq:Pi_vap}) and (\ref{eq:Pr_vap}), $T$ represents 
the temperature at the surface of accreted solids, 
for the calculation of ablation, and the temperature of the ambient 
gas for the calculation of rain out.
Equation~(\ref{eq:Pi_vap}) is taken from \citet{iaroslavitz2007}, whereas
Equation~(\ref{eq:Pr_vap}) is based on a numerical fit to data from 
\citet{melosh2007}.
Above the critical temperature, $647\,\mathrm{K}$ and $5000\,\mathrm{K}$
for water and rock, respectively, $P_{\mathrm{vap}}$ is set to infinity.
Water vapor and rock vapor are treated independently. 
Although these equations and assumptions are approximate, we do not 
expect our main results to be sensitive to their deficiencies 
at very high $T$.

\subsection{Equation of State}
\label{sec:EOS}

The equation of state of the gaseous envelope is obtained from
\citet{saumon1995} for the case of solar-abundance gas, which
applies in the outer part of the envelope. The composition of this
gas assumes $X=0.71$ (mass fraction of H), $Y=0.273$ (mass fraction of
He), and $Z=0.017$ (mass fraction of all other elements). 
For hydrogen/helium gas mixed with rock and/or water vapor, tables are 
obtained based on the quotidian equation of state of \citet{more1988}, 
as extended by \citet{vazan2013}. Separate tables exist for water vapor 
and silicate vapor. In each case, sub-tables contain the heavy material 
mixed with various mass fractions of hydrogen/helium gas, with fractions 
ranging from $0$ to $1$. If silicates and water are both present at 
a given layer, the equation of state interpolates between the two tables, 
each table weighted by the mass fraction of the corresponding species. 
The inputs to the equation-of-state tables are the density,
temperature, and composition at the local position. The tables
provide pressure, specific internal energy, and
adiabatic gradient, which are needed for the solution of the
structure equations.

\subsection{Energy Transport}
\label{sec:transp}

During the principal epoch of formation (Phases~I-IV), the outer 
layers of the model planet are characterized by energy transport 
by radiation. These layers
have close to solar (nebular) composition, and the main source
of opacity is dust grains, which enter the planet along with the
nebular gas. The grains are assumed to have a size distribution
ranging from $0.005\,\mu\mathrm{m}$ to $1\,\mathrm{mm}$. 
Details on the grain opacity appear in \citet{gennaro2016} and 
are based on tables calculated by \citet{gennaro2013}. 
In the outer regions of the planet, where grains are present, 
these opacities are roughly four times as large as those of 
\citet{movshovitz2010}, who considered grain sizes 
$>1\,\mu\mathrm{m}$ and included coagulation.
As one proceeds inward in the model, the composition becomes  
richer in heavy elements, and the opacity increases rapidly; 
the references for the opacity calculation are given in 
\citet{bodenheimer2018}. 
Thus, the inner region of the planet with heavy
element mass fraction $Z \approx 1$, becomes unstable to convection
according to the Schwarzschild criterion \citep[e.g.,][]{kippenhahn1990}. 
During the final phases of planetary evolution, after accretion stops, 
the outer part of the planet, with composition close to solar,
is also convectively unstable, except for a thin
radiative layer at the outer edge. During these phases, the grains
are assumed to sink and evaporate, and the radiative opacity
is dominated by molecular sources \citep{freedman2014}.

The region with a molecular weight gradient suppresses ordinary 
convection, as discussed in \citet{bodenheimer2018}. 
It is possible (though not mandatory) that ``semiconvection'' 
(often referred to as double-diffusive convection) can arise because 
the diffusion of heat (mediated by electrons) can be much faster 
than the diffusion of molecular species. Such mixing would tend 
to smooth out a compositional gradient and make the core more dilute 
as in \citet{leconte2012}. However, this kind of convection is driven 
by heat from below only (most of Jupiter's luminosity is driven 
by cooling from above), and such heat flow is necessarily small since 
that region is of small mass and low specific heat (hence low heat 
content, despite its high temperature). This is a very different 
situation from that discussed by \citet{leconte2012}, where 
they assumed substantial dispersal of heavy elements 
in their initial conditions. 
It is irrelevant that this convection can operate 
post-formation for billions of years since its consequences are 
limited by the small total heat content in the central region.
Quantitative details are described in Helled at al.\ (2022). 
To give a specific example, the thermal energy released by 
a $20\,\Mearth$ core of heavy elements that has cooled by 
$30\,000\,\mathrm{K}$ is a factor of several less than 
the gravitational work that would be needed to disperse 
such a core in Jupiter, even ignoring the inefficiency of 
double diffusion. We conclude that it is reasonable to exclude 
this type of convection in the calculations. 
The issues this raises are discussed further in the concluding 
section.
 
When energy transport by convection is suppressed, the calculations 
assume transport at a much reduced rate by radiation. The appropriate 
temperature gradient, which determines the radiative flux, 
is very uncertain. It is parameterized here by a fraction 
(typically $90$\%) of that given by the Ledoux criterion 
for convection in the presence of a gradient in mean molecular 
weight \citep[e.g., Equations~(4) and (5) of][]{bodenheimer2018}. 
The suppression of convection would tend to trap heat in the region 
of the planet interior to the composition gradient, and lead 
to a situation where the rate of energy deposition by the incoming
planetesimals is considerably greater than the energy radiated 
by the planet.
However, the significance of this effect depends on where 
the planetesimals break up and deposit their energy relative 
to the location of the composition gradient. 
If the energy is deposited primarily exterior to the gradient,
there is practically no effect.  

\subsection{Assumed Parameters}
\label{sec:par}

The parameters of the calculation are similar to those used in 
\citet{pollack1996} and \citet{lissauer2009}. Jupiter forms at
$a_{p}=5.2\,\mathrm{AU}$ in a nebula whose initial solid surface density 
is $\sigma^{0}_{Z}=10\,\mathrm{g\,cm}^{-2}$ at heliocentric distance
$a=a_{p}$.
The rock/ice planetesimals all have a radius of $100\,\mathrm{km}$,
and they accrete on to the planet at the rate given by 
\citeauthor{safronov1972}'s (\citeyear{safronov1972})
equation \citep[e.g.,][Equation~(2)]{bodenheimer2018},
applying the gravitational enhancement factor of \citet{greenzweig1992}.
The assumption of a fixed size is of course a crude approximation
and doubtless has some effect on the results, though the tendency
for the atmosphere to supersaturate and rain out the excess heavy
elements should occur irrespective of this assumption.

During Phases I-III of formation, the  outer boundary condition on
the planet assumes a nebular temperature of $116\,\mathrm{K}$ and
a density of $10^{-10}\,\mathrm{g\,cm}^{-3}$. The outer radius $R_p$
of the planet is set, for almost all of the formation phase, to 
$R_{\mathrm{eff}} = R_{\mathrm{H}}/4$, where $R_{\mathrm{H}}$ is 
the planet's Hill radius, as determined by three-dimensional 
hydrodynamic simulations \citep{lissauer2009}.
If the Bondi radius $R_{\mathrm{B}}= GM_{p}/c^2$, where $c$ is 
the sound speed in the disk (at $a_{p}$) and $M_{p}$ is 
the planet mass, is smaller than $R_{\mathrm{H}}/4$, then 
$R_{\mathrm{eff}} = R_{\mathrm{B}}$. 
During Phases~I through III, the gas accretion rate $\dot{M}_{XY}$ 
is determined by the requirement that $R_{p} = R_{\mathrm{eff}}$
(so $R_{p}$ increases as $M_{p}$ grows).
However, once disk-limited accretion (Phase~IV) is reached, $R_{p}$ 
contracts within $R_{\mathrm{eff}}$ and the planet ``detaches'' 
from the disk.
The temperature and density at $R_{p}$ are determined, approximately, 
by the procedure of \citet{bodenheimer2000}, which takes into account 
the shock that forms at the outer boundary of the hydrostatic planet.  
Gas and solid accretion rates during this phase are calculated
according to the method described in \citet{gennaro2021}, and
additional details are provided below.

The planet does not migrate through the disk and reaches Jupiter's 
mass ($M_{\mathrm{J}}\approx 318\,\Mearth$) at the time of the dissipation 
of the disk at Jupiter's orbital radius, $\approx 3.5 \times 10^6$~years 
in the model, or at the end of Phase V, if applicable. 
The gas dissipation time depends on the applied disk model 
during Phase~IV.

During the ensuing evolution phase, the planet contracts and cools 
as a quasi-isolated object, apart from accreting planetesimals at 
the rate of $10^{-7}\,\Mearth\,\mathrm{yr}^{-1}$ for the first ten 
million years. This additional accretion is added to our model in
order to account for collisions with small solid bodies that continued
to approach the planet subsequent to the dissipation of the gaseous
disk and to attempt to explain the observed heavy-element
enrichment of the atmosphere. Heating from the central Sun is included
in the radiative boundary condition according to Equations~(2) through
(5) of \citet{gennaro2016}. The equilibrium temperature is set 
to either $110\,\mathrm{K}$, based upon Pioneer and Voyager~1 data
\citep{hanel2003} or $102.7\,\mathrm{K}$, based on  more recent 
Cassini data \citep{lil2018}. 
As discussed in the previous section, composition gradients in 
the planet's interior are maintained during this phase.

\subsection{Accretion of Solids During Phase~IV} 
\label{sec:SAIV}

As the planet evolves through Phase~II and III, accretion of solids
tends to deplete a region, the ``feeding zone'', of half-width 
$b R_{\mathrm{H}}$ along its orbit, so that the amount of heavy elements 
in the planet is $M_{Z}\approx 4\pi a_{p} b R_{\mathrm{H}}\sigma^{0}_{Z}$
and $b$ is between $2$ and $4$.
This estimate assumes that there is no significant re-supply of solids 
to the region and that depletion only takes place because of accretion 
on the planet.
However, other processes, such as gravitational scattering, drag forces, 
and interactions among the solids can affect the delivery of solids.

Once the planet mass $M_{p}$ exceeds $\approx 50\,\Mearth$, it is 
assumed that $M_{Z}$ grows only through the expansion of the feeding 
zone into an undepleted swarm of planetesimals.
Therefore, the accretion rate of solids is tied to the total accretion
rate of the planet
\begin{equation}
    \frac{dM_{Z}}{d t}\approx \frac{4}{3}b\pi%
    \left(\frac{a^{2}_{p}\sigma^{0}_{Z}}{M_{p}}\right)%
    \left(\frac{M_{p}}{3 M_{\sun}}\right)^{1/3}%
    \frac{d M_{p}}{d t},
    \label{eq:dMZdt}
\end{equation}
where $\dot{M}_{p}=\dot{M}_{Z}+\dot{M}_{XY}$.
The effects of nearby planets and competing embryos, which can reduce 
the supply of solids, are not taken into account. 
It should be noted that, according to Equation~(\ref{eq:dMZdt}), 
the ratio of the accretion of heavy to light elements decreases 
as $M_{p}$ increases:
\begin{equation}
    \frac{\dot{M}_{Z}}{\dot{M}_{p}}\propto M_{p}^{-2/3}.
    \label{eq:dMZdMp}
\end{equation}
The ratio is linear in the quantity $b$, which is not generally
a constant but it may depend on several factors, including planet 
mass and gas drag \citep[e.g., solids' size, see][]{gennaro2015}.

In applying Equation~(\ref{eq:dMZdt}), the rate of accretion 
$\dot{M}_{Z}$ can be connected to that calculated in the prior stages 
of evolution by using the ratio $\dot{M}_{Z}/\dot{M}_{p}$ computed 
in the model (at $M_{p}\approx 50\,\Mearth$). 

\subsection{Gas Accretion During Phase~IV} 
\label{sec:GAIV}

Observations of pre-main sequence stars indicate that
$1\,\mathrm{Myr}$-old,
$M_{\star}\approx 1\,M_{\sun}$ stars accrete gas at rates on order 
of a few to several times $10^{-8}\,M_{\sun}\,\mathrm{yr}^{-1}$, 
or $\sim 0.01\,\Mearth/\mathrm{yr}$ 
\citep[e.g.,][and references therein]{demarchi2017}.
During Phase~III, rapid contraction of the planet may lead 
to accretion rates in excess of these values.
Therefore, $\dot{M}_{XY}$ may become limited by disk supply.

In a steady-state accretion disk the rate of accretion is $3\pi\nu\Sigma$
\citep{pringle1981}, where $\Sigma$ is the surface mass density of 
the gaseous component of the protoplanetary disk. For an aged system, 
one million or more years old, 
$10^{-8}\,M_{\sun}\,\mathrm{yr}^{-1}$ would correspond to a kinematic 
viscosity $\nu$ characterized by a turbulence parameter 
$\alpha\sim 10^{-3}$.
Density perturbations caused by the planet's gravity would become
substantial for $M_{p}\gtrsim 60\,\Mearth$.
During Phase~IV, since $\dot{M}_{XY}$ is less than what envelope
contraction would dictate, the planet detaches from the disk, quickly
evolving to the point where $R_p$ is much smaller than $R_{\rm eff}$,
where $R_p$ refers to the shock layer bounding the hydrostatic planet.

The accretion rate during Phase~IV depends on both disk evolution and
tidal interactions between the planet and the surrounding gas. To model
these processes, we follow the approach described in \citet{gennaro2021}.
The disk evolution is driven by viscous diffusion, photo-evaporation 
(i.e., winds from the surface), and accretion on the planet.
The disk lifetime is mainly determined by the initial surface density
of the gas, the turbulence parameter $\alpha$, and the photo-evaporation
rate (here assumed to be a constant,
$10^{-8}\,M_{\sun}\,\mathrm{yr}^{-1}$).
Applying the gas surface density at the planet's location, obtained
from the disk evolution calculation, the limiting gas accretion rate 
of the planet as a function of $\alpha$ and $M_{p}$ is derived from
three-dimensional, high-resolution hydrodynamics calculations of 
disk-planet interactions \citep{lissauer2009,bodenheimer2013}.
The accretion rates of solids and gas as a function of $M_{p}$ are
delivered to the planet formation code in the form of a table.
Phase~IV starts when $\dot{M}_{XY}$ equals the limiting rate. 
At the beginning of this phase, $dM_{XY}/dM_{p}$ may still be positive
(typically if the tidal gap is not too deep), but in these models
$\dot{M}_{XY}$ declines as $M_{p}$ increases 
\citep[see, e.g.,][]{lissauer2009,bodenheimer2013}.
Both beginning and end of Phase~IV depend on disk evolution. 

For the determination of the initial surface density of the gas,
$\Sigma^{0}$, these disk models do not assume that planetesimals
formed \emph{in situ}, i.e., that the ratio $\sigma^{0}_{Z}/\Sigma^{0}$ 
around the planet's orbit is equal to the initial dust-to-gas mass 
ratio at that location. Rather, they assume that planetesimals 
assembled from smaller bodies (formed out of dust) as they drifted 
inward from a range of distances in the disk.

\subsection{Accretion of Solids During Phase~V} 
\label{sec:SAV}

As the impact of the fragments Comet D/Shoemaker-Levy~9 into 
Jupiter's atmosphere in 1994 demonstrated, Jupiter continues to accrete 
solid material at the present epoch. 
The current accretion rate is exceedingly small, but solids were likely
accreted at a much larger rate early in Solar System history, even after 
gas accretion ceased. This has not been accounted for in our previous 
Jupiter formation simulations because solids were assumed to sink to 
the condensed core, slightly increasing the planet's $M_{Z}$ but not 
changing gas accretion, the computation of which was the focus of those 
studies.  

Here we are also concerned with the distribution of materials within 
the planet. The detailed calculation of solids accretion rates subsequent
to the dispersal of the gaseous disk around the planet is beyond the scope 
of this work. Nonetheless, as this process is likely to have operated, 
we prescribed the accretion of an additional $1\,\Mearth$ of planetesimals 
in the $10\,\mathrm{Myr}$ following the cessation of gas accretion to 
qualitatively account for  this late addition of heavy elements. 

Late accretion of solid bodies may have been the cause of the atmospheric 
enrichment.
Alternatively, enrichment might be explained by convective transport 
from deeper regions. This possibility is not consistent with our models. 
It is also discussed and discarded by Helled et al.\ (2022),
but it must be acknowledged that our understanding of convection 
in the presence of a compositional gradient is imperfect. 
These uncertainties could affect the evolution through the isolation 
phase which follows the cessation of accretion, but not significantly 
on the formation processes in Phases I through V, as detailed in the 
next section.

\section{Results} 
\label{sec:results}

Our formation models lead to a region that is almost pure heavy elements
($Z \approx 1$) near the center; that region consists of two parts.
The ``inner'' core is small and represents material that arrived
essentially unaltered because it never underwent disruption and 
only marginally vaporized in its passage through the overlying gas. 
The composition is $50$\% SiO$_{2}$ and $50$\% H$_{2}$O.
The ``outer'' core is more massive and consists of material of about 
the same composition
that arrived as droplets because the overlying gas became supersaturated
in those constituents. It is effectively $Z=1$ although it can (and does)
contain small amounts of hydrogen and helium.
By construction (and in reality) it cannot have a sharp upper 
boundary since there is no phase separation of this heavy material 
from the adjacent ``gas'', which is hot, partially degenerate metallic
fluid. 

In the following, the ``inner'' core, of mass $M_{c}$, refers to 
that part of the planet represented by an equation of state, which 
is used to determine its radius from $M_{c}$, but which is not
directly modeled. The inner core only contains heavy elements 
(although solids are $50$\% H$_{2}$O by mass, i.e., $Z\approx 0.94$, 
we assume $Z=1$ for simplicity) in the same proportions as 
accreted solids, and represents the primitive body that 
initiated planet formation; it grows as long as incoming solids 
can impact on it.

The rest of the planet (``outer'' core and above) is modeled in 
full detail.
Since the outer core lacks a sharp upper boundary, we tabulate
the ``envelope'' mass fractions including the outer core component.
We note that this issue of how to define core and envelope is pervasive
in the gravity modeling of Juno data as well and has no agreed 
resolution. Accordingly, we designate the ``envelope'', of
mass $M_{e}$, as the part of the planet surrounding the inner core
and whose structure is computed (see Section~\ref{sec:methods}).
It contains both heavy and light elements, and includes 
the ``outer'' core, defined as the envelope region
with composition $\log{Z}>-0.01$ ($Z>0.977$; H atoms in H$_{2}$O
are classified as belonging to $Z$). 
The inner plus outer cores constitute the planet's ``core''.
We indicate with $M_{e}^{Z}$ and $M_{e}^{XY}$, respectively, 
the heavy-element mass and hydrogen/helium mass of the envelope.
The total heavy-element mass of the planet is $M_{Z}=M_{c}+M_{e}^{Z}$,
whereas its total light-element mass is $M_{XY}=M_{e}^{XY}$.

%%%%%%
\begin{deluxetable*}{lccccccccccch}
%\tablenum{1}
\tablecaption{Some results from our standard model for the formation of Jupiter\label{table:smodel}
}
%\tablewidth{0pt}
\tablewidth{\textwidth}
\tablehead{
\colhead{} & \colhead{time} & 
\colhead{$M_{c}$} & \colhead{$M_{e}^{Z}$} & \colhead{$M_{e}^{XY}$} & \colhead{$T_{c}$} &
\colhead{$\rho_{c}$} & \colhead{$\log{(L/L_{\sun})}$} & 
\multicolumn{1}{c}{$\dot{M}_{Z}$} & \multicolumn{1}{c}{$\dot{M}_{XY}$} &
\colhead{$R_{p}$} & \colhead{$M_{p}$} & \nocolhead{mod} \\
\colhead{} & \colhead{(Myr)} &
\colhead{(\Mearth)} & \colhead{(\Mearth)} & \colhead{(\Mearth)} & \colhead{($10^{3}\,\mathrm{K}$)} &
\colhead{($\mathrm{g\,cm^{-3}}$)} & \colhead{} &
\multicolumn{2}{c}{($10^{-6}\,\Mearth\,\mathrm{yr}^{-1}$)} &
\colhead{($\Rearth$)} & \colhead{(\Mearth)} & \nocolhead{}
}
%\decimalcolnumbers
\startdata
Start & $0.100$ & $0.41$ & $0.00016$ & $9.7\times 10^{-6}$ & $5.70$ & $0.00087$ & $-7.12$ & $7.80$ & $0.000744$ & $61.4$ & $0.41$ &
 1725ir3e \\
Max.\ of $M_{c}$& $0.162$ & $1.26$ & $0.01$ & $0.00027$ & $14.0$ & $0.013$ & $-6.11$ & $40.0$ & $0.0656$ & $199$ & $1.27$ &
2600ir4d \\
Max.\ of $\dot{M}_{Z}$ & $0.179$ & $1.26$ & $4.43$ & $0.01$ & $25.0$ &  $0.099$ & $-5.89$ & $414$ & $3.94$ & $540$ & $5.70$ &
6200ir4de \\
End of Phase~I & $0.239$ & $1.26$ & $10.3$ & $0.46$ & $35.0$ & $0.23$ & $-7.02$ & $7.00$ & $7.00$ & $694$ & $12.0$ &
10300vin 10400ir4op \\
End of Phase~II & $2.90$ & $1.26$ & $14.84$ & $16.1$ & $46.0$ & $0.86$ & $-6.53$ & $7.02$ & $34.1$ & $927$ & $32.2$ &
21670ir4op \\
End of Phase~III & $3.01$ & $1.26$ & $17.5$ & $37.6$ & $48.0$ & $1.32$ & $-4.70$ & $125$ & $209$ & $1127$ & $56.4$ &
23800ir4opa \\
End of Phase~IV & $3.14$ & $1.26$ & $29.0$ & $286.7$ & $56.0$ & $8.55$ & $-3.80$ & $3.48$ & $0.00$ & $21.8$ & $317$ &
35150*cc7 \\
End of Phase~V & $13.3$ & $1.26$ & $30.0$ & $286.7$ & $66.0$ & $10.1$ & $-5.95$ & $0.10$ & $0.00$ & $14.5$ & $318$ &
37800*cc7 \\
Final model & $4570$ & $1.26$ & $30.0$ & $286.7$ & $57.0$ & $12.7$ & $-8.94$ & $0.00$ & $0.00$ & $10.8$ & $318$ &
41504*cc7 \\
\enddata
\tablecomments{Temperature $T_{c}$ and density
$\rho_{c}$ refer to the values at the boundary 
between inner core and envelope, that is, the inner edge of the outer core.
After $M_{c}$ achieves its maximum, planetesimals 
dissolve in the envelope.
The end of Phase~II occurs when $M_Z=M_{XY}=M_{e}^{XY}$.
Phase~IV ends when gas around the planet's orbit disperses
(i.e., $\Sigma \approx 0$).}
\end{deluxetable*}
%%%%%%

The calculation, referred to as the ``standard model'', starts with
an inner core mass of $M_{c}=0.4\,\Mearth$ with $Z=1$ and an 
envelope mass of $M_{e}=1.4 \times 10^{-4}\,\Mearth$; as stated above, 
the envelope defines the computational domain. The envelope is composed
mostly of light elements, with a small amount of heavy elements from 
the ablation of planetesimals, as they fall through the tenuous
envelope and join the inner core.
The starting time, $10^{5}$ years, is an estimate of the time required
to form a heavy-element core of that mass \citep{gennaro2014}.

\subsection{Evolution Through Phases~I and II} 
\label{sec:P12}

%%%%%%%
\begin{figure*}[t]
\centering%
\resizebox{\linewidth}{!}{\includegraphics[clip]{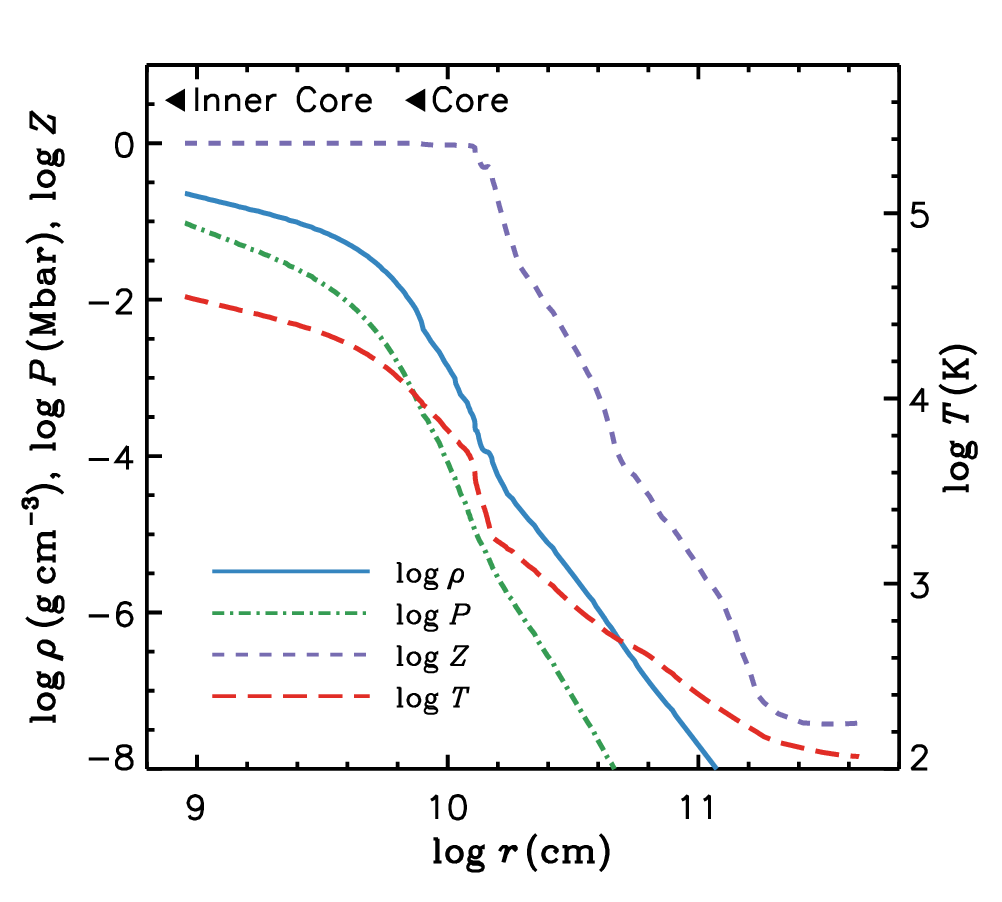}
\includegraphics[clip]{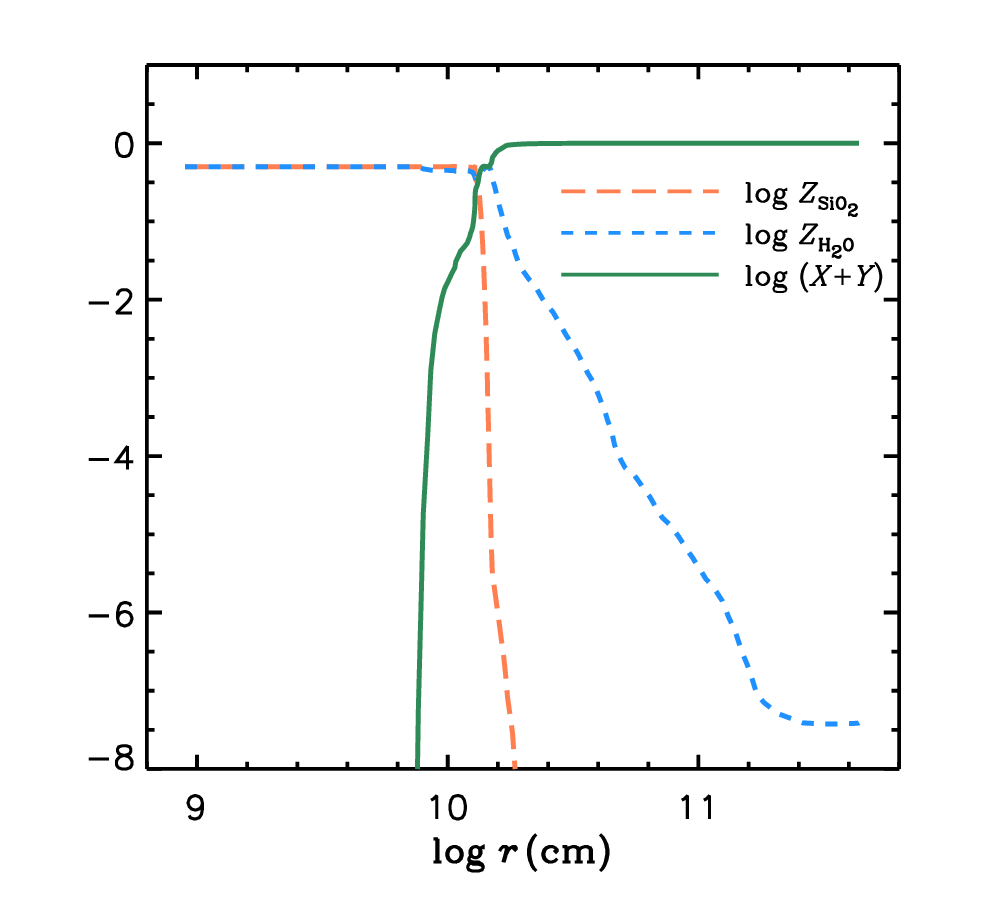}}
\caption{%
         Planet's structure and composition at 
         $t=2.43 \times 10^{5}\,\mathrm{yr}$, right after the beginning
         of Phase~II. The left panel shows density, pressure, 
         mass fraction of heavy elements, and temperature, as indicated.
         The ``inner core'' extends to the left of the curves, as marked.
         The ``core'' is the innermost region defined by $\log{Z}>-0.01$.
         The right panel shows the distributions of silicate, water and
         H-He mass fractions.
             }
\label{fig:pii}
\end{figure*}
%%%%%%%

The formation and evolution of the planet through all phases is 
summarized in Table~\ref{table:smodel}.
The columns provide the time, the inner core mass ($M_{c}$), 
the mass of heavy elements in the envelope ($M_{e}^{Z}$), 
the mass of H-He in the envelope ($M_{e}^{XY}$), 
the temperature at the inner edge of the envelope ($T_{c}$), 
the density at the same location ($\rho_{c}$), the radiated
luminosity ($L$), the mass accretion rate of planetesimals
($\dot{M}_{Z}$), the mass accretion rate of H-He ($\dot{M}_{XY}$),
the outer radius ($R_{p}$), and the total mass ($M_{p}$).
In the earliest phases the planetesimals hit the inner core.
Once the core mass reaches $M_{c} = 1.26\,\Mearth$ 
($t=1.62 \times 10^{5}\,\mathrm{yr}$; $M_{e} \approx 10^{-2}\,\Mearth$),
the planetesimals break up in the envelope and remain there 
(by assumption); the inner core mass remains constant from 
then on. For the remainder of Phase~I, breakup occurs at 
envelope temperatures large enough for H$_{2}$O and SiO$_{2}$
to evaporate. Thus, an outer core is formed with mass partitions
of the accreted solids.

During Phase~I, the accretion rate $\dot{M}_{Z}$ increases  to 
a maximum at $1.79 \times 10^{5}\,\mathrm{yr}$.
The maximum in the radiated luminosity occurs roughly contemporaneously.

There are differences in our models compared to those obtained 
in previous 
calculations \citep[see, e.g.,][]{lissauer2009} in which the solids 
are assumed to eventually sink to the core. One such difference here 
is that almost all of the added heavy-element material remains in 
the envelope. 
At the maximum in $\dot{M}_{Z}$, $M_{e}^{XY}=0.01\,\Mearth$, which 
is only a very small fraction of the total envelope mass 
$M_{e} = 4.44\,\Mearth$.
The mean molecular weights are higher than those in a H-He envelope 
and therefore, in order to maintain hydrostatic equilibrium, 
the internal temperatures and densities must be higher. The higher 
densities result in a larger capture cross section for 
the planetesimals' accretion, so the maximum 
$\dot{M}_{Z} = 4.14 \times 10^{-4}\,\Mearth\,\mathrm{yr}^{-1}$ 
is larger than that in previous calculations. 
For example, in \citet{lissauer2009}, the corresponding value is 
$\dot{M}_{Z} = 8.8 \times 10^{-5}\,\Mearth\,\mathrm{yr}^{-1}$, 
and in \citet[][their case $\sigma10$]{movshovitz2010} it is 
$\dot{M}_{Z} = 1.3 \times 10^{-4}\,\Mearth\,\mathrm{yr}^{-1}$. 
Also, the internal temperature in the \citeauthor{movshovitz2010} 
model at the same total mass ($M_{p}=5.7\,\Mearth$) is about a factor
$4$ smaller than the value given in Table~\ref{table:smodel}.
The maximum luminosity is lower by about an order of magnitude
than those in the comparison cases. This difference is caused 
in part by planetesimals not sinking to the core but breaking up 
and vaporizing near $r=1.4 \times 10^{10}\,\mathrm{cm}$, 
rather than sinking to $r\approx 1.4 \times 10^{9}\,\mathrm{cm}$ as 
in previous cases. Also, some of the accreted
energy is prevented from escaping through the composition gradient. 

The structure at $t=1.79 \times 10^{5}\,\mathrm{yr}$ consists of a 
$Z \approx 1$ region (silicates and water) from the inner core radius 
$R_{c}=9.0 \times 10^{8}\,\mathrm{cm}$ to
$r=1.49 \times 10^{10}\,\mathrm{cm}$, 
where $T=3300\,\mathrm{K}$. 
The silicate mass fraction decreases from $50$\% to $1$\% between 
that radius and
$r=1.6 \times 10^{10}\,\mathrm{cm}$ and
temperature $2600\,\mathrm{K}$. A nearly pure water layer extends out
to $r=2.9 \times 10^{10}\,\mathrm{cm}$, where $T=467\,\mathrm{K}$,
and is formed by ablation of the planetesimals. 
Between that radius and $r=3.5 \times 10^{10}\,\mathrm{cm}$, 
where $T=420\,\mathrm{K}$, the water vapor mass fraction decreases 
from $100$\% to $1$\%.
The outer radius is $R_{p} = 3.44 \times 10^{11}\,\mathrm{cm}$. 
The breakup of planetesimals occurs around 
$r = 1.40 \times 10^{10}\,\mathrm{cm}$,
at temperatures high enough so that both water and silicates can 
vaporize and mix with the ambient gas. This layer is at the outer 
edge of the region with $Z \approx 1$, so that there is some 
suppression of energy transport in the overlying region with 
a gradient in the silicate fraction.
The radiated luminosity entering that region from below is a factor 
$2.7$ larger than that exiting at the top of the region.
Farther out, the gradient in the water fraction
does not contribute to heat trapping or suppression of
convection, because it is in a radiative region.

The value of $\dot{M}_{Z}$ then steadily declines with time as the mass
of solids available to accrete onto the planet is gradually depleted. 
At $ t=2.39 \times 10^{5}\,\mathrm{yr}$, when $M_{p}=12\,\Mearth$,
$\dot{M}_{XY}$ exceeds $\dot{M}_{Z}$, signalling the end of Phase~I. 
This mass is about the same as the corresponding mass in 
\citet{lissauer2009} and \citet{movshovitz2010}, but the time is shorter
than that found by \citet{lissauer2009}, $4.3 \times 10^{5}\,\mathrm{yr}$.
A different value for $M_p$ ($7.3\,\Mearth$) was obtained by
\citet{gennaro2014}, who used a more sophisticated and much more 
computationally-intensive prescription for the calculations of $\dot{M}_{Z}$.
The planet structure is still dominated by the heavy elements,
whose total mass is $M_{Z}=11.56\,\Mearth$.
The inner region with $Z \approx 1$ extends out to $T=10^{4}\,\mathrm{K}$  
and $r=7.5 \times 10^{9}\,\mathrm{cm}$ and includes about $6\,\Mearth$.
The gradient in silicate vapor extends from there down to 
$T=2500\,\mathrm{K}$ and out to $r=1.4 \times 10^{10}\,\mathrm{cm}$. 
The gradient in water vapor extends down to $T=1000\,\mathrm{K}$, 
where $r=2.4 \times 10^{10}\,\mathrm{cm}$. Outside that point, very
little planetesimal material has been accreted.
Most of the deposition of mass and energy by planetesimals occurs 
at around $r=1.34 \times 10^{10}\,\mathrm{cm}$, about $15\,R_{c}$ 
and near the outer edge of the layer with a silicate composition 
gradient, that is, at about 1.1 times the
radius of the outer core.
Thus, there is a negligible amount of suppression of energy transfer 
because of the gradient.
The layers with a gradient in the water mass fraction also do not 
suppress energy transfer, and there is no mixing of material there, 
because energy transfer is by radiation from
$r=1.5 \times 10^{10}\,\mathrm{cm}$ out to the surface, at 
$r=4.4 \times 10^{11}\,\mathrm{cm}$. The radiated luminosity
is again a factor $10$ lower than in \citet{lissauer2009}, 
although the values of $\dot{M}_{Z}$ are similar.
The temperature at the inner core-envelope boundary is
$\approx 3.5 \times 10^{4}\,\mathrm{K}$. 
Figure~\ref{fig:pii} shows the structure variables (left) and 
the distribution of the silicate, water, and H-He constituents 
(right) somewhat after the beginning of Phase~II.

Phase~II evolves on a longer timescale. 
At $2.4 \times 10^{6}\,\mathrm{yr}$,
the masses in H-He and heavy elements are, respectively,
$M_{XY} = 7.0\,\Mearth$ and $M_{Z} = 14.0\,\Mearth$. 
At this time,
$\dot{M}_{Z}=2.5 \times 10^{-6}\,\Mearth\,\mathrm{yr}^{-1}$, and 
$\dot{M}_{XY} = 7.66 \times 10^{-6}\,\Mearth\,\mathrm{yr}^{-1}$, 
a ratio of about $3$. Accretion continues with a gradually increasing 
ratio up to $t=t_{\mathrm{cross}}=2.90 \times 10^{6}\,\mathrm{yr}$, 
when $M_{Z}=M_{XY}$, the crossover point. 
The value of the crossover mass, $M_{\mathrm{cross}}=M_{Z}=16.1\,\Mearth$, 
is almost exactly the same as that in \citet{lissauer2009} and
\citet{movshovitz2010}. The value of $t_{\mathrm{cross}}$ is
about $20$\% longer than in \citet{lissauer2009}, and much longer 
than in \citet{movshovitz2010}, who obtained a value of only 
$10^6\,\mathrm{yr}$. The short time in the latter work is most
likely a result of their calculation of grain settling and coagulation 
in the envelope, which results in a significant reduction in opacity.
Note, however, that \citet{gennaro2021} also include the grain 
settling and coagulation as well as a detailed calculation of 
the dynamics of the planetesimal swarm, with a large range of 
planetesimal sizes. They obtain $M_{\mathrm{cross}}= 9.8\,\Mearth$ 
and $t_{\mathrm{cross}} = 2.38 \times 10^6\,\mathrm{yr}$.
The sharp reduction in opacity due to grain settling and coagulation
does not occur, because the small planetesimals ($\lesssim 1\,\mathrm{km}$ 
in radius) ablate in the outer regions of the envelope
and resupply the population of small grains.

During Phase~II, the envelope density increases to the point at which
planetesimal breakup occurs where envelope temperatures are large 
enough for H$_2$O to fully evaporate and remain around the breakup layer. 
Silicates, however, do not evaporate completely and sink, forming a layer 
in which the gradients of $Z_{\mathrm{SiO_{2}}}$ and $Z_{\mathrm{H_{2}O}}$
differ. Farther out, $Z$ is dominated by H$_2$O.

In the present calculation, at $t_{\mathrm{cross}}$, the outer core 
region with $Z \approx 1$ extends to a radius 
$r=2.7 \times 10^{9}\,\mathrm{cm}$ 
and includes a mass of about $8\,\Mearth$. Farther out, $Z$ decreases 
to a value of $10$\% at $r=2.9 \times 10^{10}\,\mathrm{cm}$. 
By comparison, the core radius in \citet{lissauer2009} and
\citet{movshovitz2010} is $R_{c}=2.0 \times 10^{9}\,\mathrm{cm}$, 
bounding a mass of $16\,\Mearth$. Thus, in our case, there is 
a centrally condensed ``core'' component, but the distribution 
of heavy elements is much more extended in radius than in 
the comparison cases.
Another point of comparison is the temperature $T_{c}$ 
at the inner core-envelope boundary. 
In our case, at $r=2 \times 10^{9}\,\mathrm{cm}$ ($0.75$ times 
the radius of the outer core), 
$T=3.3 \times 10^{4}\,\mathrm{K}$, while in the comparison 
cases, at the same radius, 
$T_{c} \approx 1.85 \times 10^{4}\,\mathrm{K}$.

\subsection{Evolution Through Phases~III and IV} 
\label{sec:P34}

%%%%%%%
\begin{figure*}[t]
\centering%
\resizebox{\linewidth}{!}{\includegraphics[clip]{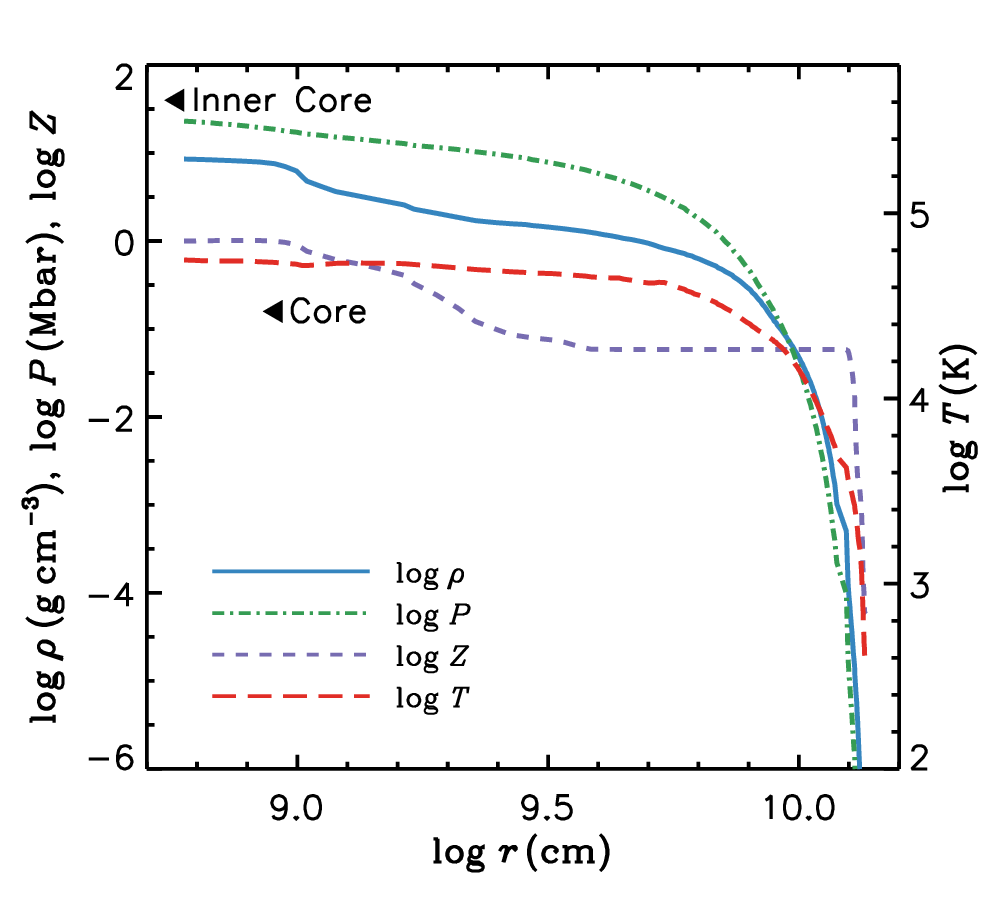}
\includegraphics[clip]{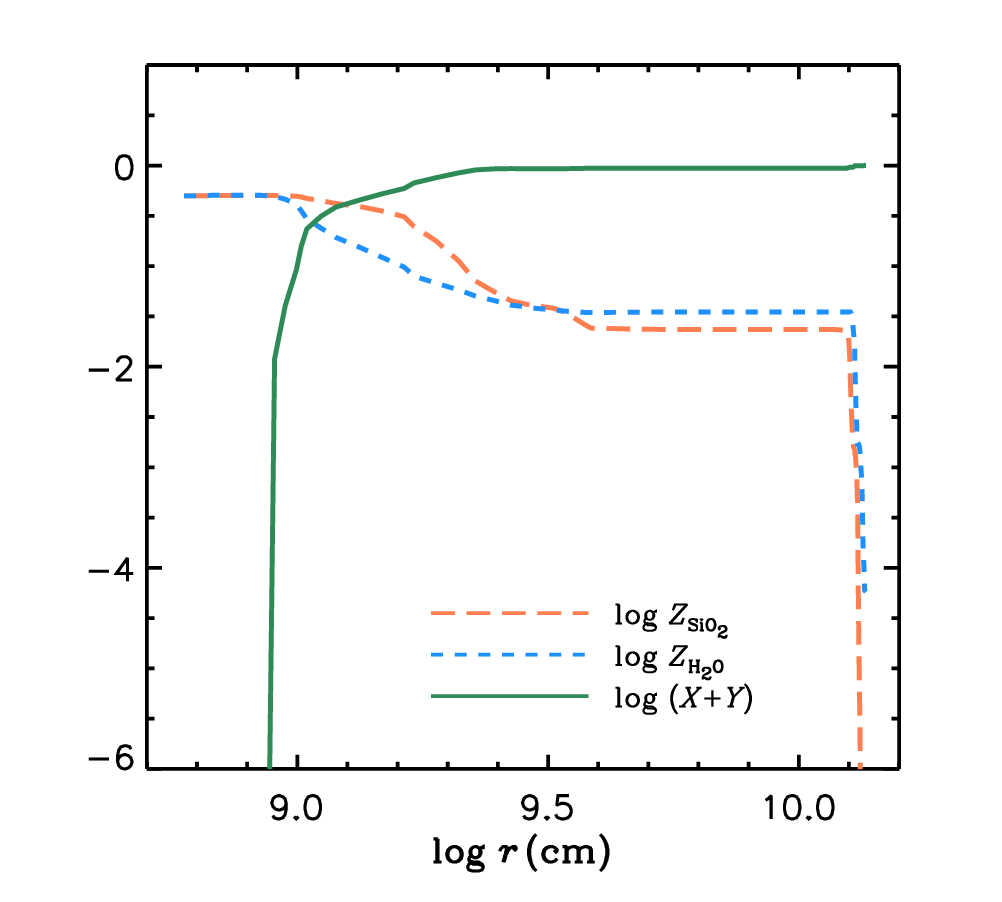}}
\caption{%
         Planet's structure and composition at the end of Phase~IV, 
         $t=3.14 \times 10^{6}\,\mathrm{yr}$. 
         The structure variables ($P$, $\rho$, $T$) and the heavy-element 
         mass fraction ($Z$) are displayed on the left, and the interior
         distributions of silicates, water, and H-He are shown on the right.
         The core region (as defined herein) is also indicated.
}
\label{fig:piv}
\end{figure*}
%%%%%%%

Crossover ($M_{Z}=M_{XY}$)
marks the end of Phase~II. During Phase~III, the value of
$\dot{M}_{XY}$ increases 
by a factor of $56.6$, while the rate $\dot{M}_Z$ also increases 
significantly, by a factor of $17.8$ (see Table~\ref{table:smodel}).
The mass $M_{Z}$ grows from $2.7\,\Mearth$ to $18.8\,\Mearth$, 
while $M_{e}^{XY}$ grows from $21\,\Mearth$ to $37.6\,\Mearth$.
This phase includes what is generally referred to as  the 
``runaway accretion'' phase, which is defined by a decreasing 
growth timescale, $M_{p}/\dot{M}_{p}$, as the planet mass increases, 
e.g., $\dot{M}_{p}\propto M^{\beta}_{p}$ with $\beta>1$.
In the standard model discussed herein, proper runaway accretion
starts after Phase~III begins and ends together with Phase~III.
In the model, the planet acquires about ten percent of its final 
mass during runaway accretion.

For  an assumed disk viscosity parameter 
$\alpha = 4 \times 10^{-3}$, $\dot{M}_{XY}$ exceeds the maximum 
rate at which the disk can supply gas to the planet at 
$t=3.01 \times 10^6\,\mathrm{yr}$, when $M_{p}=56.4\,\Mearth$. 
At this time, the planetesimals break up at about 
$r=3.3 \times 10^{10}\,\mathrm{cm}$, at the outer edge of 
the layer with the composition gradient.
The radiated luminosity, $\log{(L/L_{\sun})} = -4.7$, arises 
primarily from gravitational contraction rather than from energy 
deposition by planetesimals.  The planet 
has reached its maximum size, with 
$R_{p}= 7.2 \times 10^{11}\,\mathrm{cm}$. 

In Phase~IV, consistent with the assumption regarding $\dot{M}_Z$ 
during Phases~I--III, the parameter $b$ (see Section~\ref{sec:SAIV}) 
is set to $4$. It refers to the half-width, in Hill radii, of the 
region from which the solids can accrete onto the planet. 
The value of $\dot{M}_Z=\dot{M}_Z(t)$ is obtained as described in
that section. The value of $\dot{M}_{XY}$ versus time is discussed 
in Section~\ref{sec:GAIV}.
Gas accretion increases somewhat in the beginning but it then decreases 
with time because of gap opening by tidal torques, as also illustrated
in Figure~2 of \citet{bodenheimer2013}, and gas dispersal.
About $80$\% of the planet's final mass is acquired during this phase, 
in a non-runaway fashion.
The planet contracts rapidly as soon as this phase sets in; 
at the maximum of $\dot{M}_{XY}$, $R_{p}=280\,\Rearth$. Temperatures
at $R_p$, which refers to the layer just interior to the shock, depend
on $R_p$, $M_p$, and $\dot{M}_{XY}$ and are typically a few hundred 
kelvins.
The radiated luminosity during this phase consists of two components:
the internal luminosity $L_{\mathrm{int}}$ primarily from contraction,
and the accretion luminosity 
$L_{\mathrm{acc}} \approx G M_p \dot{M}_{XY}/R_{p}$, arising from 
the infall of gas onto the planet.
The maximum of 
$L = L_{\mathrm{int}} + L_{\mathrm{acc}}$ occurs at about 
$t=3.05 \times 10^{6}\,\mathrm{yr}$, when $M_{p}=236\,\Mearth$, 
$R_{p} = 30.8\,\Rearth$, and 
$\dot{M}_{XY}=2.02 \times 10^{-3}\,\Mearth\,\mathrm{yr}^{-1}$.
At maximum power output, $\log{(L/L_{\sun})} = -3.2$, 
of which $\approx 22$\% is contributed by $L_{\mathrm{int}}$.
Phase~IV ends at $t=3.14 \times 10^{6}\,\mathrm{yr}$ with 
$M_{Z}= 30.26\,\Mearth$, $M_{e}^{XY} = 286.7\,\Mearth$, and 
$M_{p}=317\,\Mearth$.
At this time, the accretion rate of heavy elements is
$\dot{M}_{Z} = 3.48 \times 10^{-6}\,\Mearth\,\mathrm{yr}^{-1}$.
As mentioned in Section~\ref{sec:SAIV}, the rate of solids accretion 
declines relative to  $\dot{M}_{XY}$ during this phase, from about 
$9$\% at the beginning to about $2.8$\% toward the end.

The structure of the planet ($M_{p}=317\,\Mearth$) at the end of 
gas accretion is shown in Figure~\ref{fig:piv} (left panel), which 
gives, as a function of radius, the density, pressure, temperature,
and composition.
The core region has a mass of several times \Mearth. 
Outside that region, the composition gradient extends out to 
$r=0.27\,R_{p}$ and 
a temperature of $4.56 \times 10^{4}\,\mathrm{K}$. 
The energy transport is by ordinary convection beyond that radius, 
and the composition is uniform with $Z \approx 0.06$ in the region
with water and silicates, and $Z\approx 0.04$ in the overlying region
with just water.
Note that the mass fraction of water exceeds that of silicates
in the uniform region, while the silicates dominate in the gradient 
region (see Figure~\ref{fig:piv}, right). 
The total mass of silicates equals that of water throughout 
the evolution (because of the assumed composition of accreted solids).
The convection zone extends out to 
$r\approx 1.3 \times 10^{10}\,\mathrm{cm}$,
where the temperature is $2000\,\mathrm{K}$.
Beyond this point, the silicate vapor condenses and is removed
via rain-out. 
In the outer layers, energy transport is by radiation.
In the very outermost layers, the water vapor mass fraction drops off
not because of condensation ($ T \approx 400$ K), but because little
material landed there by ablation during Phase~IV.
The total luminosity of $\log{(L/L_{\sun})} = -3.8$
comes primarily from the accretion luminosity of the last remaining
infalling gas onto the planet, which is re-radiated from behind
the accretion shock. The internal luminosity, generated primarily
from contraction, is $\log{L_{\mathrm{int}}/L_{\sun}} = -5.7$.

\subsection{Phase V}
\label{sec:P5}

%%%%%%%
\begin{figure*}
\centering%
\resizebox{\linewidth}{!}{\includegraphics[clip]{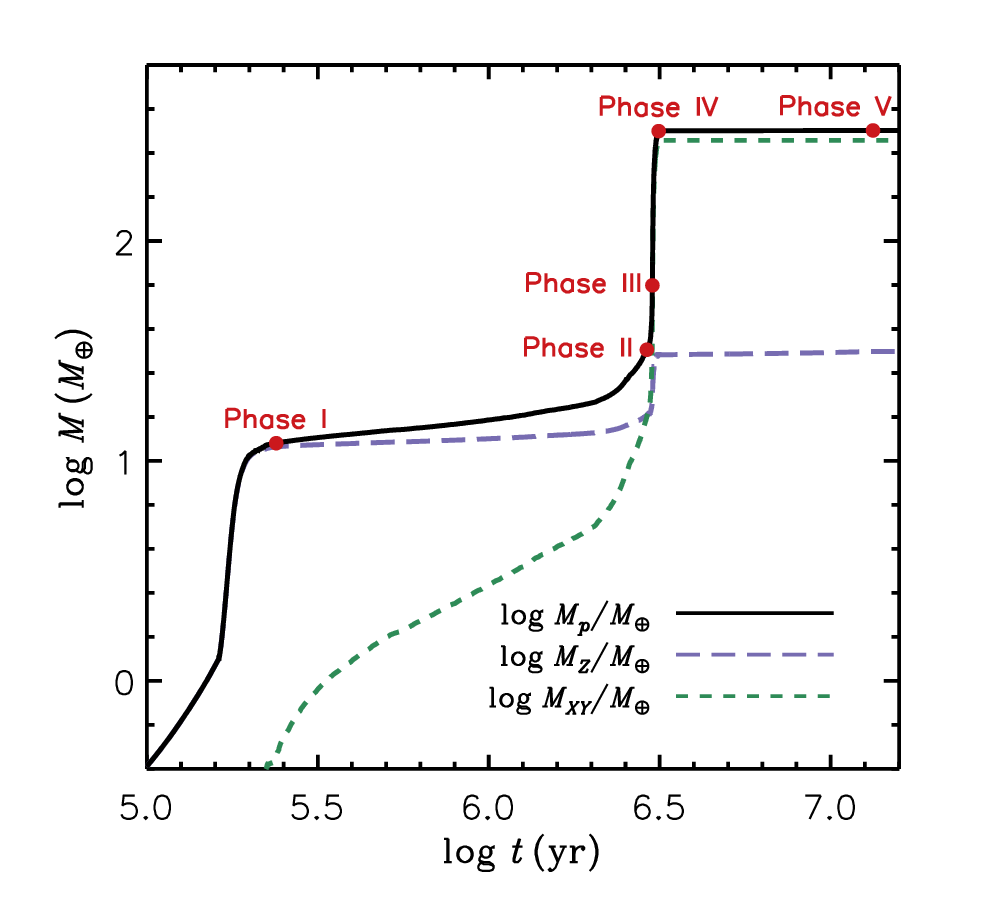}
\includegraphics[clip]{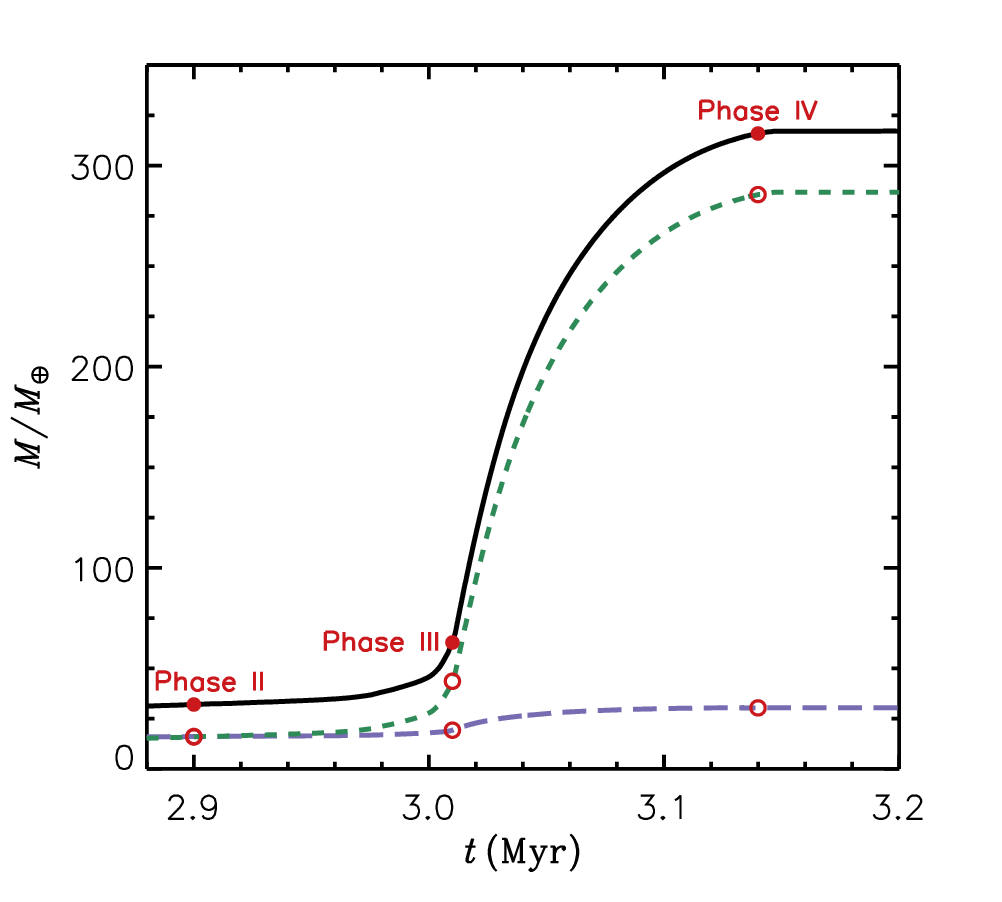}}
\caption{%
         Left:
         Planet growth as a function of time, throughout formation.
         The curves represent total, heavy-element and H-He mass, 
         as indicated in the legend.
         The five solid circles indicate the end of Phases~I through V, 
         as labelled.
         Right:
         Same quantities as in the left panel, but masses versus time 
         are illustrated on linear scales during Phases~III and IV only.
         Solid and open circles indicate the end of the phases on each 
         curve. 
         The gas accretion is smooth across the transition between 
         Phases~II and III ($M_{XY}=M_{Z}$), and increases during
         Phase~III. $\dot{M}_{XY}$ declines as the planet mass increases
         during Phase~IV.
             }
\label{fig:mass}
\end{figure*}
%%%%%%%

At the beginning of Phase~V the planet radius 
$R_{p} \approx 2\,R_{\mathrm{J}}$, 
where $R_{\mathrm{J}}=7 \times 10^{9}\,\mathrm{cm}$.
Accretion of solids from residual planetesimals is assumed to continue
after gas accretion stops, at an arbitrary rate of 
$10^{-7}\,\Mearth\,\mathrm{yr}^{-1}$ for a time of $10^{7}$ years.
The final mass is $M_{p}=318\,\Mearth$, of which $31.26\,\Mearth$
is heavy elements. At this time, $t=1.33 \times 10^{7}\,\mathrm{yr}$,
the contraction has led to slight adiabatic heating of the inner regions; 
later on the cooling phase sets in. The outer region of the planet is
low $Z$ (i.e., primarily H and He); thus the added heavy-element material 
produces a Rayleigh-Taylor instability.
Mixing of material results in uniform composition $Z \approx 0.06$
in the region from the outer edge of the composition gradient out
to where the silicates condense, at $T \approx 3300\,\mathrm{K}$.
Outside that point, a region of water vapor with $Z \approx 0.037$ 
exists out to the surface, at $T \approx 520\,\mathrm{K}$.
Figure~\ref{fig:mass} shows the temporal evolution of $M_{Z}$, $M_{XY}$,
and $M_{p}$, up to just beyond this point in time (left panel).
The positions at which the transition between two Phases occurs are 
marked by circles.
The right panel of the Figure shows the mass evolution during Phases~III 
and IV. At the transition between Phases~II and III, when $M_{XY}=M_{Z}$,
the gas accretion rate does not vary sensibly, and increases significantly
only later in Phase~III. The H-He mass curve also indicates that
$\dot{M}_{XY}$ decreases during Phase~IV, as the planet acquires 
most of its mass.

\subsection{Post-formation Evolution} 
\label{sec:Pi}

%%%%%%%
\begin{figure}
\centering%
\resizebox{\linewidth}{!}{\includegraphics[clip]{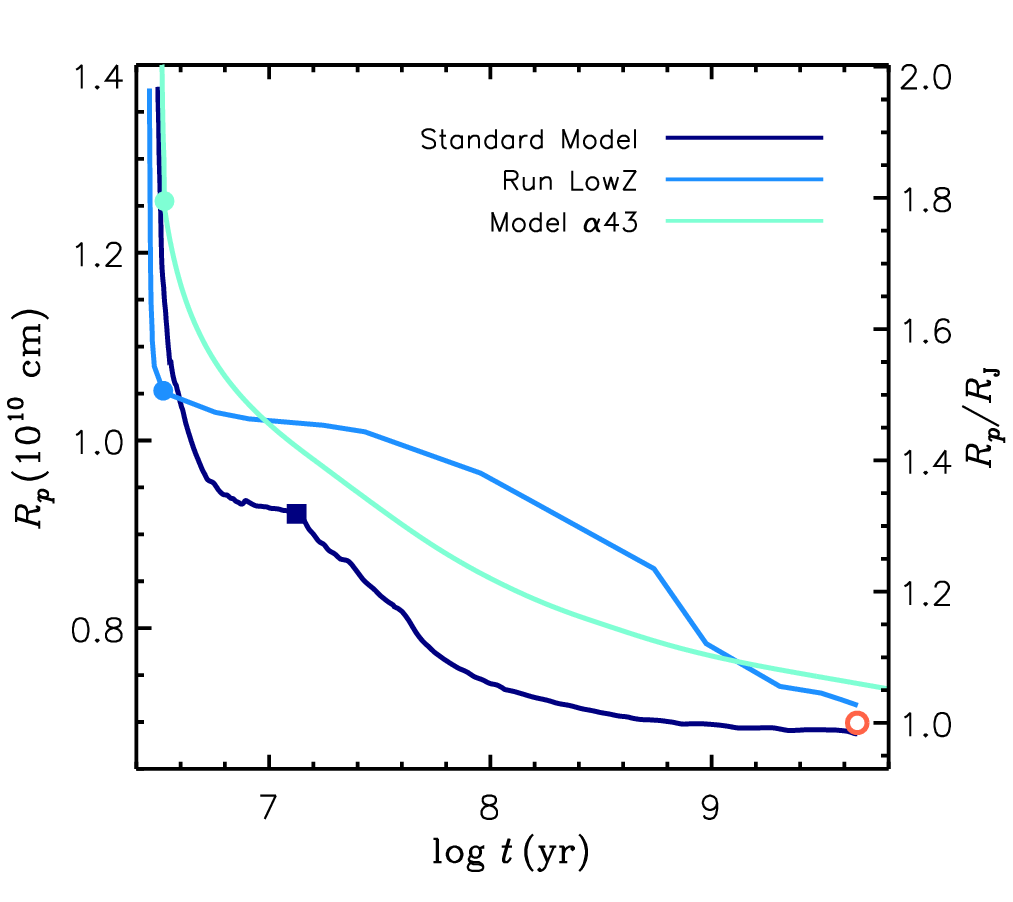}}
\caption{%
         Planet radius, $R_{p}$, during the isolation phase and
         up to the current age, obtained from our standard model
         (as indicated). The planet's volume contracts nearly $8$ times 
         during this period. The final radius differs by less than 
         $2$\% from Jupiter's radius, $R_{\mathrm{J}}$ (open circle).
         Also plotted is the radius obtained from Run LowZ, with 
         lower final $M_{Z}$, discussed in Section~\ref{sec:var},
         and from one of the models of \citet{gennaro2021}, in which
         heavy elements sediment to the core.
         Solid symbols mark the end of formation: circles for
         the end of Phase~IV and square for the end of Phase~V.
         }
\label{fig:rp}
\end{figure}
%%%%%%%

%%%%%%%
\begin{figure}
\centering%
\resizebox{\linewidth}{!}{\includegraphics[clip]{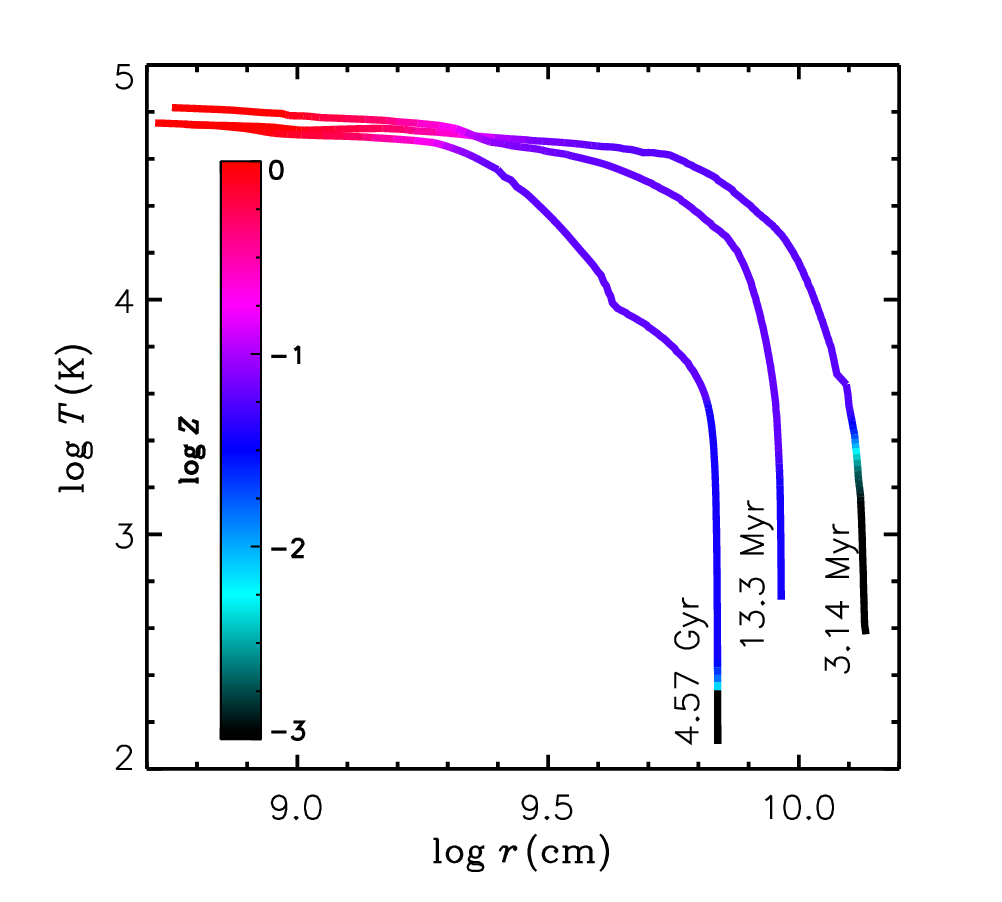}}
\caption{%
         Temperature distribution at three different epochs
         of evolution: end of Phase~IV ($3.14\times 10^{6}\,\mathrm{yr}$),
         end of Phase~V ($13.3\times 10^{6}\,\mathrm{yr}$),
         and current age. The color bar renders the logarithm of
         the heavy-element mass fraction. Uniform-composition
         layers correspond to portions of the curves of uniform
         colors. Note that the temperatures $T_{c}$ at the inner edge 
         of the outer core are comparable 
         at $t= 3.14 \times 10^{6}\,\mathrm{yr}$ and 
         $t=4.57 \times 10^{9}\,\mathrm{yr}$ 
         (see Table~\ref{table:smodel}).
             }
\label{fig:t3}
\end{figure}
%%%%%%%

Further evolution involves no additional accretion. The value of $R_{p}$
declines, as shown in Figure~\ref{fig:rp}, ending at 
$t= 4.57 \times 10^{9}\,\mathrm{yr}$ (the age of the solar system) 
about $1.5$\% below the actual (volumetric mean) radius $R_{\mathrm{J}}$.
For comparison, the Figure also reports the radius obtained from another
model with final $M_{Z}=23\,\Mearth$, discussed in Section~\ref{sec:var},
and from a model in which the heavy elements sink to a compressible core
\citep[see][]{gennaro2021}.
The internal luminosity (neglecting the re-radiated solar energy) declines 
by three orders of magnitude, to end at
$\log{L_{\mathrm{int}}/L_{\sun}} = -8.94$,
in very good accord with recent measurements of this quantity, $-8.92$
\citep{lil2018}. 
In the outer layers of the final model, at a pressure of 
$1\,\mathrm{bar}$, 
the temperature is $160\,\mathrm{K}$. Note that the standard adiabat 
for calculation of Jupiter interior models starts at a temperature of 
$165\,\mathrm{K}$ at $1\,\mathrm{bar}$ pressure \citep[e.g.,][]{stevenson2020}.
The temperature distribution agrees with data within a few percent,
down to $20\,\mathrm{bar}$ \citep{seiff1998}.
These results are modestly affected by the assumed value of 
the equilibrium temperature, which depends on the measured albedo. 
The temperature as a function of radius in the models at three 
different epochs is  shown in Figure~\ref{fig:t3}, in which
uniform colors on the curves represent envelope layers of uniform 
composition.
%%%%%%%
\begin{figure*}
\centering%
\resizebox{\linewidth}{!}{\includegraphics[clip]{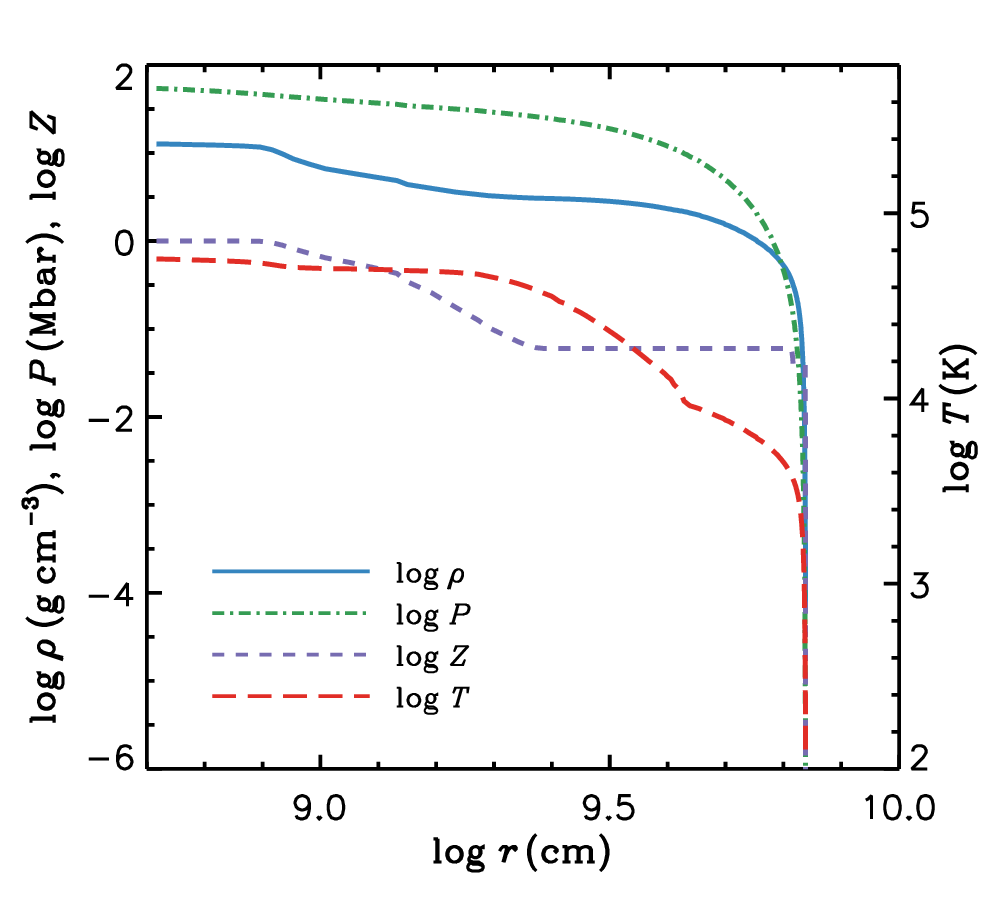}
\includegraphics[clip]{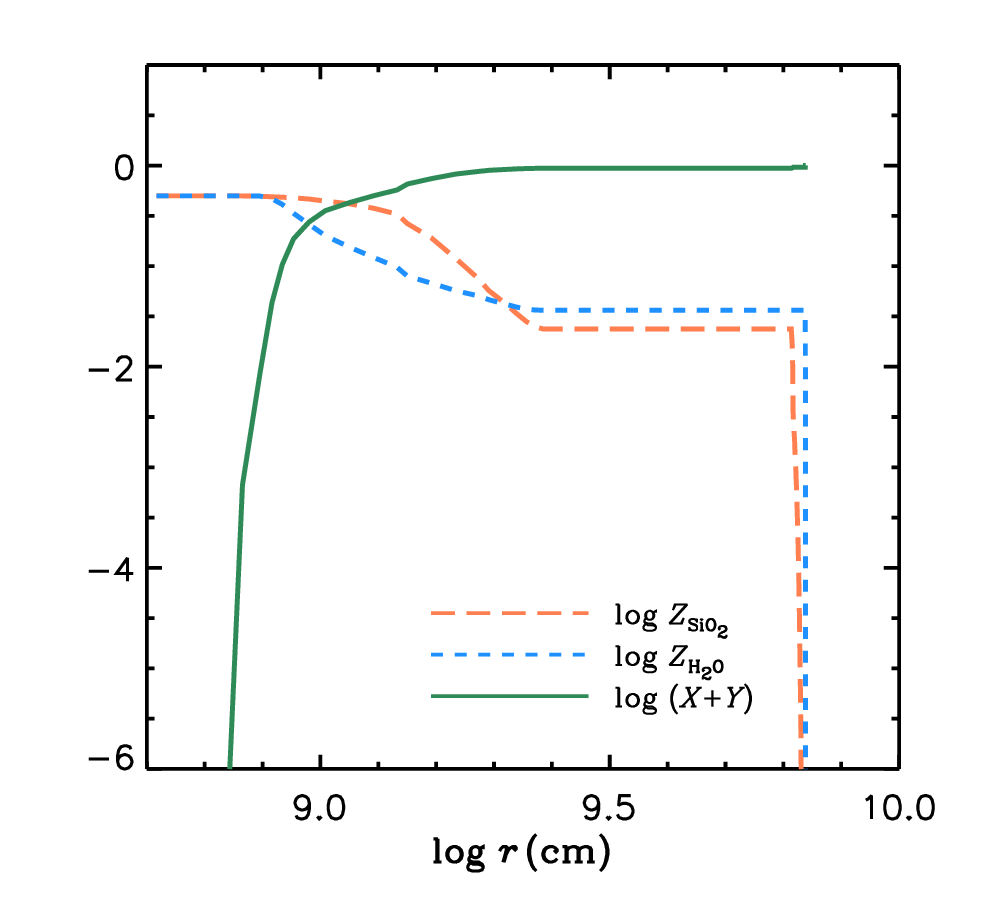}}
\caption{%
         As in Figures~\ref{fig:pii} and \ref{fig:piv}, 
         but the interior structure (left)
         and composition (right) refer to the current age, 
         $t= 4.57 \times 10^{9}\,\mathrm{yr}$.
             }
\label{fig:cage}
\end{figure*}
%%%%%%%
Figure~\ref{fig:cage} shows the structure at the current age,
in the same format as in Figure~\ref{fig:piv}.
The pressure at the inner core-envelope interface is $54\,\mathrm{Mbar}$.
The outer core region with $Z>0.98$ extends to 
$r=7.9 \times 10^{8}\,\mathrm{cm}$
and includes a mass of about $5\,\Mearth$. The layers with a gradient
in composition extend out to $r=2.3 \times 10^{9}\,\mathrm{cm}$,
about $33$\% of the planet's radius, $R_{p}$.
The relative mass fractions of silicates and water, as a function
of radius, exhibit a similar behavior to that shown 
in Figure~\ref{fig:piv}.
From that radius out to $r=6.53 \times 10^{9}\,\mathrm{cm}$
the composition is uniform in both silicates and water with
$Z_{\mathrm{SiO_{2}}}+Z_{\mathrm{H_{2}O}}=0.06$.
Beyond that point the silicates condense, and the water
layer has uniform $Z_{\mathrm{H_{2}O}}=0.037$ out to 
$r= 6.88 \times 10^{9}\,\mathrm{cm}$ and $T=271\,\mathrm{K}$.
In the very outer layers, where the temperature drops to about
$125\,\mathrm{K}$, the water condenses and rains out. The region of
uniform composition, between $r=2.3 \times 10^{9}\,\mathrm{cm}$
and $r=6.88 \times 10^{9}\,\mathrm{cm}$, is marked by a uniform color
($\log{Z} \approx -1.22$) in Figure~\ref{fig:t3}.

Figure~\ref{fig:piv} (right) and Figure~\ref{fig:cage} (right) display 
the effect of the very different evaporation temperatures for ice and 
rock: the rock is deeper and the ice is shallower. 
This “differentiation” is not directly due to density (except that 
it is gravitationally stable) but to the thermodynamics of vaporization. 
It implies that the relative abundances of rock and ice in the region
towards the surface (yet deeper than current cloud formation) should 
be different from what was originally accreted, with the rock being 
less abundant. Unfortunately, this important prediction is difficult 
to test with current models of the giant planets. It may however be 
a very important conclusion of relevance to the so-called ice giants 
such as Uranus and Neptune, or exoplanets (sub-Neptunes), since this 
effect is then more likely to be expressed in the heat flow and perhaps
even the gravity field.

\subsection{Simulations with Alternative Prescriptions for Late-Stage Solids Accretion} 
\label{sec:var}

A modification to this calculation (Run NoFive) was made in which 
no solid accretion was postulated after the end of gas accretion 
($t=3.14 \times 10^{6}\,\mathrm{yr}$), i.e., no Phase~V.
The evolution up to this time is the same as given 
in Table~\ref{table:smodel}.
The model beyond this point has total heavy-element mass 
of $30.26\,\Mearth$ out of the total mass of $318\,\Mearth$. 
At the final time of $4.57 \times 10^{9}\,\mathrm{yr}$, the radius 
is less than half a percent larger than $R_{\mathrm{J}}$, and 
the internal luminosity is $\log{L_{\mathrm{int}}/L_{\sun}} = -8.95$, 
very close to that of the standard model.
The temperature at $1\,\mathrm{bar}$ pressure again is $160\,\mathrm{K}$.
The values of $T_{c}$ and $\rho_{c}$ are, respectively,
$6.3 \times 10^{4}\,\mathrm{K}$ and $11.9\,\mathrm{g\,cm^{-3}}$.
The water vapor abundance in the outer layers (above the condensation
temperature) is $Z_{\mathrm{H_{2}O}}=0.0347$, only slightly lower than
that in the standard case. The reason for this small difference
is that the $1\,\Mearth$ of added heavy elements in the standard model
is mixed through the outer convective envelope, which includes
$\approx 280$ times that mass. 
Furthermore, the silicate component of added material is supersaturated 
and rains out to a layer where the temperature is $\approx 3000\,\mathrm{K}$.
To make a significant difference in the $Z$ abundance of the surface 
layers, say $1$\%,  $\approx 3\,\Mearth$ would have to be added 
during Phase~V. 
It remains to be determined whether such a process is feasible. 

A second modification (Run LowZ) tests the effects of the parameter
$b$ in Equation~(\ref{eq:dMZdt}), the half-width of the region from 
which the planet can collect solid material. The value of $b=4$ 
\citep[taken from][]{kary1994}
used above is changed to $b\approx 2.3$, the value derived from 
detailed calculations of the evolution and accretion of 
the planetesimal swarm surrounding the forming planet
\citep{gennaro2014,gennaro2021}.
Details on the procedure for the calculation of $\dot{M}_{Z}$ and
$\dot{M}_{XY}$ during Phase~IV are reported in the latter paper.
The rates are based on a disk model whose initial mass is about
$0.1\,M_{\sun}$ and whose initial gas surface density is about
$700\,\mathrm{g\,cm^{-2}}$, consistent with the assumption used
in the present calculations.
The viscosity parameter, as above, is $\alpha= 4 \times 10^{-3}$.

%%%%%%%
\begin{figure*}
\centering%
\resizebox{\linewidth}{!}{\includegraphics[clip]{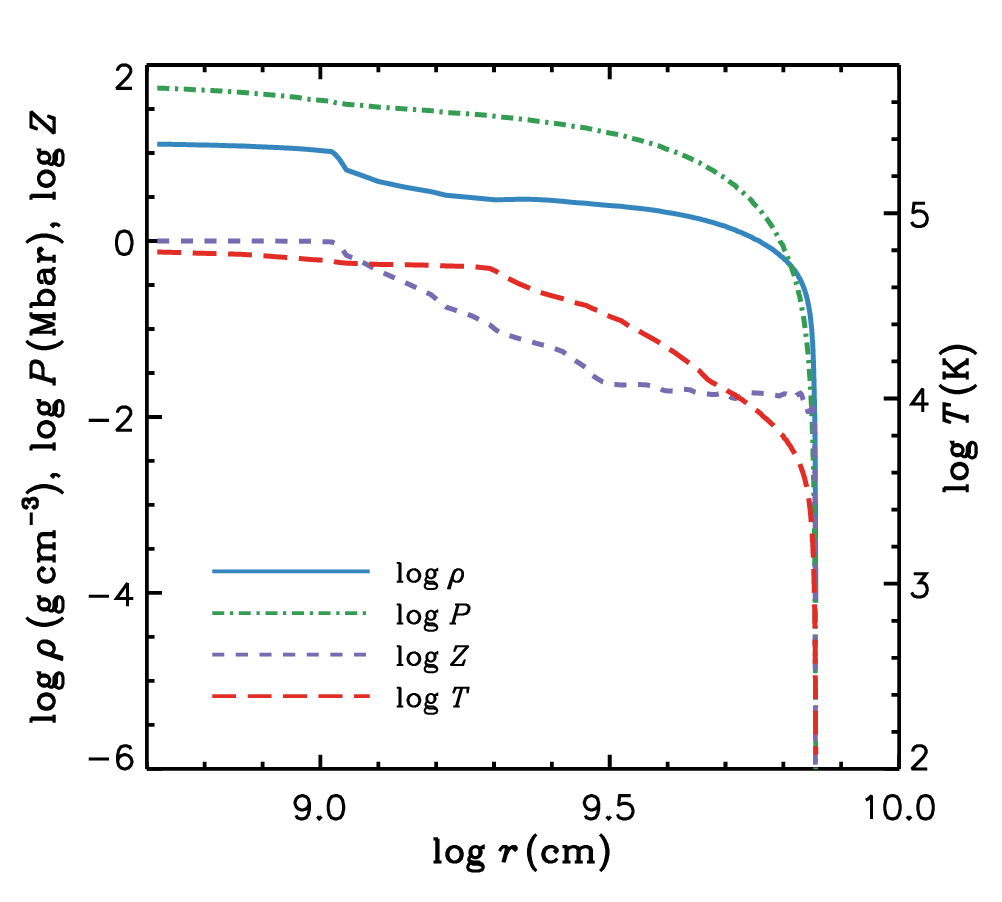}
\includegraphics[clip]{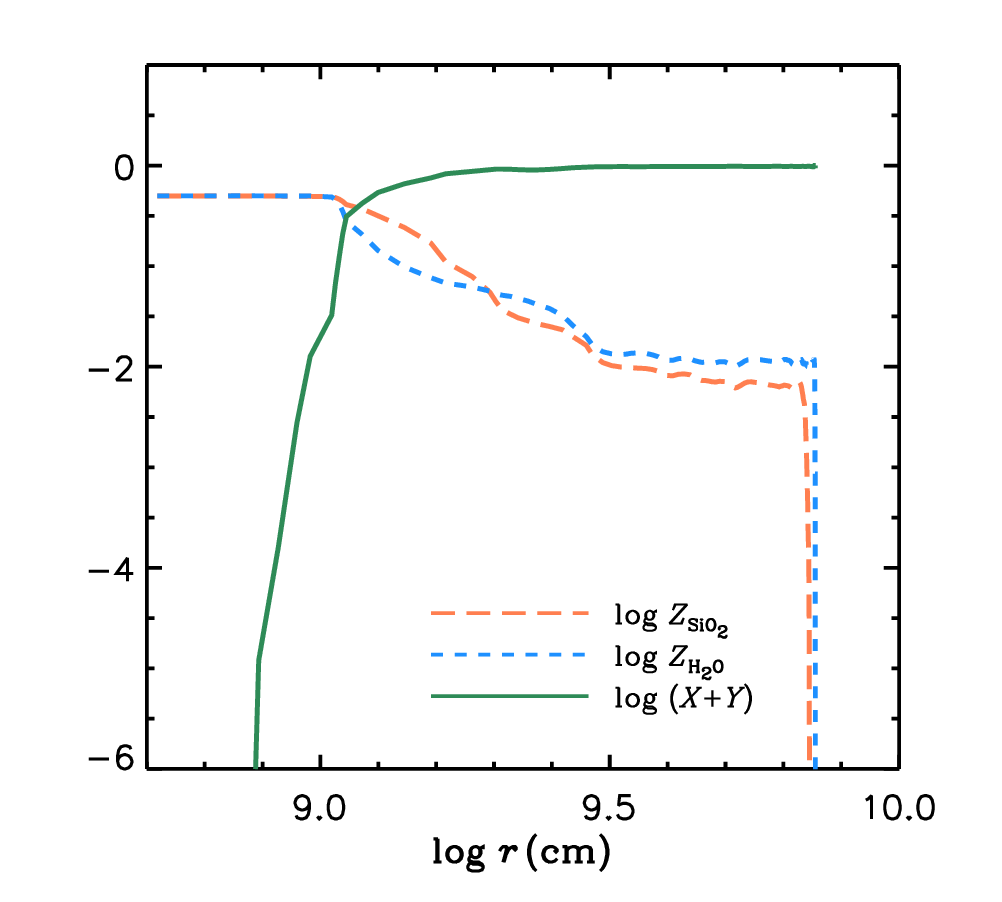}}
\caption{%
         Planet's structure and composition at the current age
         for Run LowZ. Compare with the standard model, illustrated
         in Figure~\ref{fig:cage} at the same epoch. The left panel
         shows the interior structure while the right panel
         shows the distributions of mass fractions of light and
         heavy elements.
             }
\label{fig:cage_2}
\end{figure*}
%%%%%%%
The evolution up to the onset of disk-limited accretion is the same
as that given in Table~\ref{table:smodel}. The gas accretion rate
during Phase~IV is the same as above, with the same maximum rate,
but the values for $\dot{M}_{Z}$ are lower than before.
The end of accretion of gas and solids occurs at 
$t=3.32 \times 10^{6}\,\mathrm{yr}$,
with $M_{c} = 1.26$, $M_{e}^{Z} = 21.77$, $M_{e}^{XY} = 294.6$,
and $M_{p} = 317.6$, all in units of \Mearth. The value of $M_{Z}$
at the end of accretion is $23\,\Mearth$, considerably lower than
the value of $31.26\,\Mearth$ obtained in the standard model.
At this time, $T_{c}=7.26 \times 10^{4}\,\mathrm{K}$,
$\rho_{c} = 8.85\,\mathrm{g\,cm^{-3}}$,
$\log{(L/L_{\sun})}  = -6.95$, and 
$R_{p}=17.4\,\Rearth = 1.6\,R_{\mathrm{J}}$.
Shortly before the end of gas accretion, the effective temperature 
(defined by $L_{\mathrm{acc}} + L_{\mathrm{int}} =
4 \pi \sigma_B R_p^2 T_{\mathrm{eff}}^4$) is $\approx 1100\,\mathrm{K}$,
compared with $\approx 1440\,\mathrm{K}$ in the standard case. 
 
The evolution into the isolation phase has no added accretion
of solids (no Phase~V), reaching $t= 4.57 \times 10^{9}\,\mathrm{yr}$ 
with $R_{p}$ about $2.5$\% higher than $R_{\mathrm{J}}$,
$\log{L_{\mathrm{int}}/L_{\sun}} = -9.4$, 
$T_{c}= 6.2 \times 10^{4}\,\mathrm{K}$
and $\rho_{c}=12.6\,\mathrm{g\,cm^{-3}}$. The central core
with $Z \approx 1$ includes $6.2\,\Mearth$, and the composition gradient
extends out to $r=3.185 \times 10^9\,\mathrm{cm}$, about $44$\% of 
$R_{p}$.
The deep uniform-composition region of the envelope, with 
$Z \approx 0.022$, extends out to $T\approx 3000\,\mathrm{K}$, where 
the silicates start to condense and rain out.
Farther out, the water layer has $Z_{\mathrm{H_{2}O}} \approx 0.016$,
except in the very outer layers where water condenses. These values
are lower than in the standard model. The structure and composition are
illustrated in Figure~\ref{fig:cage_2}.

%%%%%%%
\begin{figure}
\centering%
\resizebox{\linewidth}{!}{\includegraphics[clip]{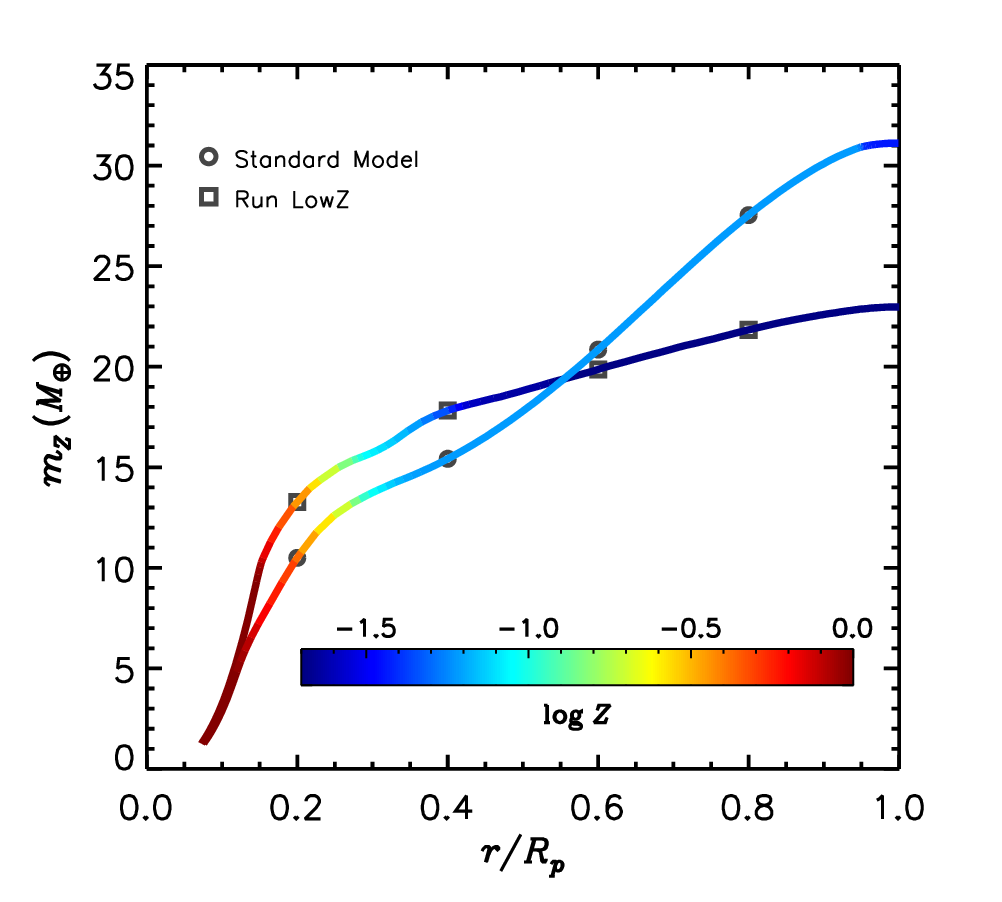}}
\caption{%
         Integrated mass of heavy elements in the planet interior,
         at the current age, for the standard model and the 
         Run LowZ, with lower $M_{Z}$, and color rendering of $\log{Z}$.
             }
\label{fig:mz}
\end{figure}
%%%%%%%

The distribution of heavy elements in Run LowZ is compared to 
that of the standard model, at the current epoch, in Figure~\ref{fig:mz}.
Note that the two models have comparable cores ($Z>0.98$) and both
have $\approx 20\,\Mearth$ of heavy elements within half of the planet
radius.

\section{Discussion} 
\label{sec:disc}

The interior structures of the three cases presented here 
(Standard, NoFive and LowZ), after $4.6\,\mathrm{Gyr}$, are
similar. All have a central core ($Z \approx 1$) where the radius of
the outer core ranges from $7.5 \times 10^{8}\,\mathrm{cm}$ to
$9.5 \times 10^{8}\,\mathrm{cm}$ ($11$\% to $13$\% of the planet radius
$R_{p}$), enclosing masses between $4.5$ and $6.2\,\Mearth$.
Outside this region, a composition gradient exists, with $dZ/dr < 1$,
extending out to radii between $33$\% to $44$\% of $R_{p}$.
Outside the gradient layer, the composition is nearly uniform in
the convective envelope. In this region the values of $Z$ range from 
$2.2$ to $6.0$\%. The values of $Z$ in the H$_{2}$O layer near
the outer edge of the envelope, where the silicates have condensed, 
range from $1.6$ to $3.7$\%.

\subsection{Previous Formation and Evolution Models}

\citet{ormel2021} calculated models through Phase~II, based
on accretion and dissolution of dust and pebbles in the H-He 
atmosphere; they also found a gradient in composition.
In calculations at $5\,\mathrm{AU}$, with particles in 
the size range $0.01$--$1\,\mathrm{cm}$, the somewhat 
diluted core has $Z$ values in the range $0.5$ to $1.0$.

A model similar to that considered herein,
in which Jupiter's evolution is followed through
the isolation phase, starting from an assumed heavy-element distribution, 
is presented by \citet[][their case $J_{2}$]{vazan2016}
under simpler physical assumptions. At the current age, $Z$
decreases from $1$ at the center to $0.05$ at mass
fraction $0.3$ (radius fraction $\approx 0.5$); farther out $Z$ is constant.
Convection is suppressed in the gradient layer. Similar calculations through 
the isolation phase are reported by \citet{vazan2018}.

\citet{lozovsky2017} calculated the deposition of heavy elements in
the envelope during the formation phase. At the end of Phase~IV, the mass
fraction of heavy elements decreases from $1$ at the center to about $0.1$
at mass fraction $0.25$. They also find, as we do, that interior 
temperatures are much higher than temperatures would be for a homogeneous, 
isentropic H-He envelope. 
Starting with assumed distributions of $Z$ at the onset of Phase~IV,
(motivated by \citeauthor{lozovsky2017} results) 
\citet{mueller2020} 
perform calculations up to the present day. The results at the final time 
show that the region enriched in $Z$ extends to no more than mass fraction 
$0.2$ ($40$\% of the radius). 
Note that our present calculations make no assumption regarding the initial 
distribution $Z(r)$ other than that accretion starts with a heavy-element 
core ($Z=1$) of $0.4\,\Mearth$.

An alternative calculation of the full formation and evolution of Jupiter
has been published by \citet{gennaro2014,gennaro2021}.
In that calculation, the accretion rate 
of solids is based on a detailed statistical calculation of the 
evolution and accretion of a planetesimal swarm surrounding the
forming planet, with a considerable range of planetesimal sizes. The
present calculations do not include this (very time-consuming) effect, 
but use the simpler accretion rates of \citet{greenzweig1992}
with a single planetesimal radius of $100\,\mathrm{km}$. 
The typical planetesimal size  that provides the bulk of 
the heavy-element mass accreted in the 
\citeauthor{gennaro2021} calculations is around $50\,\mathrm{km}$. 
Another effect included in 
those calculations is the coagulation and settling of the small grains, 
which enter the planet along with the nebular gas and those released 
by ablating planetesimals, and the resulting effects on the opacity 
in the outer radiative zone that is found during formation. 
The present calculations do not include this effect, but instead use 
a fixed opacity table as a function of temperature, density, and 
composition.
The computed opacities in radiative zones in our calculations and 
in the calculation by \citet{gennaro2014,gennaro2021} are not markedly
different \citep[][their Figure~11]{gennaro2014}. 
The major effect that is not included in the \citeauthor{gennaro2021} 
calculations is the deposition of the heavy elements within the planet,
and the resulting change in envelope composition.
The \citeauthor{gennaro2021} calculation assumes that all accreted heavy 
elements sink to the core.
The final core mass turns out to be about $20\,\Mearth$. 
It is planned in the future to improve the present calculations to
account for the solid accretion rates of \citet{gennaro2014,gennaro2021},
including a range of planetesimal sizes.

\subsection{Observationally-Constrained Models of Jupiter's Interior}

There are still small but significant unresolved differences
in the modeling efforts to explain gravity data obtained
by the Juno mission.
One such effort, led by B.~Militzer, follows from the pioneering 
effort of \citet[][]{wahl2017}. Another, led by Y.~Miguel,
follows from the pioneering work of \citet[][]{guillot2005}.
Interior models are inevitably non-unique because they are 
designed to fit a small number of observable quantities 
(e.g., radius, $J_{2}$, $J_{4}$, and $J_{6}$, which may be affected 
by differential rotation).
As a consequence, some features of interior models such as a
$Z \approx 1$ core may be indeterminate if it is small enough
(e.g., a few $\Mearth$ at most).
Interior modeling also requires good understanding of the equation 
of state of hydrogen, and errors of a percent or two in 
density at a specified pressure are possible and can affect 
estimates of $Z(r)$.
Some of Jupiter's gravity-based models include the effect of 
helium immiscibility interior to $85$\% of $R_{\mathrm{J}}$ 
and a change in $Z$ abundance at that point. This effect 
was not included in our calculations, as it is not likely 
to be of importance to the issues we seek to address.
Nonetheless, differences in the gravity-based models are probably
smaller than the differences between those models and our formation
models.

The key difference is that our formation models tend to produce
a much more centrally concentrated distribution of heavy elements,
even though they no longer have a sharp boundary between a central
core, with $Z=1$, and an overlying envelope, with $Z\ll 1$.
The main aspects of this difference can be appreciated by describing
some published interior models that fit the Juno data.

The preferred model from \citet{debras2019} has the following layers:
\textit{i}) a possible $Z=1$ central core out to $5$\% of the radius,
\textit{ii}) a diluted core with $Z\approx 0.15$ at $5$\% of the radius,
decreasing  to $Z \approx 0.005$ at $65$ to $70$\% of the radius;
\textit{iii}) a convection zone with uniform $Z \approx 0.005$ out
to $80$-$90$\% of the radius,
\textit{iv}) a layer in which $Z$ increases outward to the surface
value of $0.02$, and
\textit{v}) an outer convection zone with uniform $Z=0.02$.
Thus, the dilute core is considerably more extended in radius than
the dilute region in our models.
A central, $Z=1$ core is permitted in their models but not required 
by gravity data (F.~Debras, 2021, personal communication).

The work of \citet{wahl2017} considers models with a dilute core 
of constant $Z$ extending outward from the center to $15$--$60$\% 
of $R_{\mathrm{J}}$. The cores contain $7$ to $25\,\Mearth$ 
of heavy elements and $Z$ is relatively low; thus the core region 
consists mainly of H-He. Outside the core there are two layers
with different $Z$, both significantly smaller than the core value.
This structure is not consistent with the physical assumptions and
results of our models. It requires post-formation erosion of 
the primitive $Z=1$ region and upward mixing of the heavy elements 
in the H-He gas. Alternatively, a much larger ratio 
$\dot{M}_{XY}/\dot{M}_{Z}$ is required during Phases~I and II, 
which seems difficult to achieve within the core accretion scenario.

Both \citet{debras2019} and \citet{wahl2017} (as well as more recent work)
have difficulties matching the threefold enhancement, relative to solar,
of the heavy elements in the envelope's outer layers, as measured by 
the Galileo probe. Since we have no reason to doubt the probe results 
(or older estimates of methane enrichment in the atmosphere), this 
discrepancy raises the possibility of equation of state errors. 
It is not known whether such an error would only affect the determination 
of $Z$ in the outer region or might even affect the issue of deep structure
that is our focus here.
Helled et al.\ (2022) discuss the discrepancies between gravity-based 
models and the present interior models based on formation and evolution. 
The possible sources of this discrepancy are enumerated at the end 
of the summary below.

\section{Summary and Conclusions} 
\label{sec:conclusions}

This paper describes calculations that follow the formation and 
evolution of Jupiter, starting from a small core of heavy elements, 
continuing through the phases of concurrent accretion of solids 
(planetesimals) and nebular gas, and ending at the age of the solar 
system. The calculations presented herein depart from the historical 
assumption that all accreted heavy elements sink to a condensed 
core and that all the gas outside the core has nebular 
composition. Here we calculate the dissolution of the planetesimals, 
composed of silicates and ice, in the gaseous envelope. 
The end result is a central region of $\approx 100$\% heavy elements 
with a mass of several $\Mearth$, outside of which is a layer 
of steadily decreasing mass fraction of heavy elements. 
The layer with a composition gradient ($dZ/dr\neq 0$) 
extends outward to $33$--$44$\% of the planet radius 
($ \approx R_{\mathrm{J}}$); 
above it is a layer of nearly uniform composition with a few percent 
of heavy elements by mass. The central region and the outer layers 
with nearly uniform composition are separate convection zones with 
different specific entropy.

The calculations are self-consistent in the sense that once 
the distribution of heavy elements in the envelope is determined at 
a given time step, the resulting structure is obtained through 
a full solution of the planetary structure equations, taking into account 
the effects of the heavy elements on the equation of state and opacity. 
The resulting structure is hot enough that most of the accreted solids 
vaporize and mix with H-He. However, this flux of accreted
material is high enough that the gas tends to oversaturate, with 
the excess heavy elements sinking as droplets toward a region, 
which may be supercritical.
The region with the composition gradient is stable against
ordinary convection, so that transfer of energy and material through
this region is strongly suppressed. This effect, along with that 
of the mean molecular weight in the equation of state, results in 
a considerably hotter interior than that found in most previous
calculations. The temperature gradient in the region with the composition
gradient is uncertain and must be considered in a more detailed manner
in the future. 
Because of considerably different evaporation temperatures, silicates
and H$_{2}$O have quite different radial distributions through most 
of the formation phases.

In the standard model presented herein, the accretion era of the planet
separates into five phases. Phase~I is dominated by accretion of solids, 
much of which forms a $Z \approx 1$ core. The phase differs from that 
in previous calculations \citep[e.g.,][]{lissauer2009,movshovitz2010} 
in that $\dot{M}_{Z}$ is noticeably higher, the luminosity 
is lower, and the duration of the phase is shorter 
(see Section~\ref{sec:P12}).
Phase~I ends when $\dot{M}_{XY} = \dot{M}_{Z}$, 
with $M_{p} = 12\,\Mearth$ at $t=2.4 \times 10^5$ years. 
Phase~II takes an order of magnitude longer than Phase~I and 
is characterized by relatively low but increasing $\dot{M}_{XY}$
and $\dot{M}_{Z}$.
The ratio $\dot{M}_{XY}/\dot{M}_{Z} \approx 3$ increases with time. 
This Phase ends when $M_{XY}=M_{Z}$ (crossover), 
at $t=2.9 \times 10^{6}\,\mathrm{yr}$ and $M_{Z}=16\,\Mearth$, 
the same as in \citet{lissauer2009} and \citet{movshovitz2010}. 
The structure at this time is different, however, with higher internal 
temperatures and with about half of $M_{Z}$ in a $Z \approx 1$ core, 
the remaining portion residing in a region with a composition gradient, 
extending out to ten core radii. 

Phase~III, with  $M_{XY} > M_{Z}$, starts at crossover and ends at 
$t=3.01 \times 10^{6}\,\mathrm{yr}$. The gas accretion rate increases
rapidly, to $\dot{M}_{XY} = 2 \times 10^{-3}\,\Mearth\,\mathrm{yr}^{-1}$. 
At this stage, the contraction of the planet is so rapid that the demand 
on $\dot{M}_{XY}$ to satisfy the boundary condition
$R_{p}=R_{\mathrm{eff}}$ cannot be satisfied by the disk, signalling 
the start of Phase~IV, when $M_p \approx 56 \Mearth$.

The planet then begins to contract, and the boundary condition shifts 
from nebular values to shock conditions at the boundary between 
the infalling gas and the hydrostatic planet. 
The disk-limited gas accretion rates are based on three-dimensional 
hydrodynamic simulations of a planet embedded in a disk 
\citep{lissauer2009,bodenheimer2013}. 
The solids still accrete during this phase, but at a much lower rate 
than the gas. The value of $\dot{M}_{Z}$ is somewhat uncertain, 
as it depends on the size of the feeding zone, from whose edges 
the solids are assumed to originate. 
Two cases are considered in this paper. The first (standard) case
assumes that the half-width of the feeding zone is $4\,R_{\mathrm{H}}$. 
At the end of Phase~IV, the total heavy-element mass is $M_{Z}=30.3\,\Mearth$.
In the second case (Run LowZ), based on the detailed calculations
of the evolution of the planetesimal swarm \citep{gennaro2021},
the half-width is $2.3\,R_{\mathrm{H}}$ and the final $M_{Z}=23\,\Mearth$.
Phase~IV ends at the time of disk dissipation,
$t=3.14 \times 10^{6}\,\mathrm{yr}$ in the standard case and 
$t=3.3 \times 10^{6}\,\mathrm{yr}$ in Run LowZ.  

The new Phase~V is introduced in the standard case. It has been 
hypothesized that the enhanced abundances of heavy elements, compared 
to solar values, in Jupiter's outer layers can be caused by 
the addition of solid material to the planet after gas accretion ends.
During the first $10^{7}$ years following disk dispersal, the planet 
is assumed to accrete $1\,\Mearth$ of planetesimals. As the planet has 
contracted substantially and the scale height in the upper envelope is 
far smaller than during earlier phases, these planetesimals 
break up high in the envelope, locally adding heavy elements.
The high density in those layers results in a Rayleigh-Taylor 
instability and mixing of $Z$ material into the outer convection zone. 
The $1\,\Mearth$ is diluted in the entire mass of that zone, 
$\approx 280\,\Mearth$. The water abundance in the outer 
layers hardly differs from the case with no added solids
(i.e., Run NoFive). Considerably more $Z$ material would
have to be added to make a difference.
However, the total $M_{Z}$ added to the planet in Phase~IV 
does have an effect on the water abundance in the outer layers, 
which turns out to be about $3.7$\% in the standard case 
($M_{Z}=31.3\,\Mearth$) versus $1.6$\% in the alternate case 
(Run LowZ, $M_{Z}=23\,\Mearth$).

As mentioned at the beginning of this section, the final models at 
$t=4.57 \times 10^{9}\,\mathrm{yr}$ have a central (inner plus outer)
core with $Z \approx 1$ 
that has less mass than in previous calculations in which all added 
solid material sinks to the core. Additional $Z$ material is spread 
out in a layer with a gradient in composition, $dZ/dr < 0$, extending 
out to at most $44$\% of $R_{p}  \approx R_{\mathrm{J}}$.
Models of present Jupiter \citep[e.g.,][]{wahl2017,debras2019},
which provide reasonable fits to the gravitational moments determined 
by Juno \citep{iess2018}, typically have no core with $Z=1$ but, 
instead, regions with low $Z$ extending out to large radius fractions.
The formation pathway for such a structure is unclear. Our formation 
models do not provide a solution to this dilemma, but it should be 
pointed out that models with a central $Z \approx 1$ core of a few 
$\Mearth$, which fit Juno data, are not ruled out
\citep{guillot2018,debras2019}. 
Current modeling effort led by Y.~Miguel \citep{miguel2022}
and B.~Militzer (2022, personal
communication) also suggest that a central core of a few Earth masses 
is possible.

These formation models fail to explain the extent of dilution 
heavy-element suggested by interior models based on gravity data. 
There are four possible reasons for this discrepancy:
\textit{i}) The accretion model is wrong;
\textit{ii}) The equation of state is wrong (i.e., current static 
models that fit gravity are wrong);
\textit{iii}) one or more giant impacts occurred to stir up 
the core; 
\textit{iv}) Post-formation mixing by convection created 
the dilute core. 
All of these have been discussed above but we here offer a summary.
Concerning the first, it is possible that the delivery of solids 
was different than in our picture. Concerning the second, 
the inability to explain the current atmosphere suggests problems, 
and the current thermodynamic descriptions being used disagree with 
each other at a level that is small but could change the models, 
suggesting the need for more laboratory work. Concerning the third, 
giant impacts are not yet adequately modeled, and the one published 
example showing extensive upward mixing of a core was based on 
a head-on collision that is unlikely relative to multiple oblique 
collisions. 
Concerning the fourth, there is currently no sufficiently high 
resolution, low viscosity convection simulation that determines 
the consequences of Reynolds stress (turbulence, effectively) 
in upward mixing arising from cooling from above (the dominant 
mechanism for Jupiter's luminosity). All four possible explanations 
suggest that there is much scope for future work in several directions.

\acknowledgments

We thank Kevin Zahnle, Richard Freedman, and two anonymous reviewers 
for insightful comments that helped improve the paper. 
Primary support for this work was provided by NASA's 
Emerging Worlds program, proposal 18-EW18\_2\_0060.
GD acknowledges support from NASA ROSES grant 80HQTR19T0071.
PB acknowledges support from NASA Origins grant NNX14AG92G. 
A significant contribution was made by R.~Helled, who provided 
the equation-of-state tables for mixtures including silicates and water.
Computational resources supporting this work were provided by the NASA 
High-End Computing (HEC) Program through the NASA Advanced 
Supercomputing Division at Ames Research Center.

%\vspace{5mm}
%\facilities{HST(STIS), Swift(XRT and UVOT), AAVSO, CTIO:1.3m,CTIO:1.5m,CXO}

%\software{astropy \citep{2013A&A...558A..33A},  
%          Cloudy \citep{2013RMxAA..49..137F}, 
%          SExtractor \citep{1996A&AS..117..393B}
%          }

%\bibliography{loc}{}
%\bibliographystyle{aasjournal}

%% This command is needed to show the entire author+affiliation list when
%% the collaboration and author truncation commands are used.  It has to
%% go at the end of the manuscript.
%\allauthors

%% Include this line if you are using the \added, \replaced, \deleted
%% commands to see a summary list of all changes at the end of the article.
%\listofchanges

\end{document}